\documentclass[conference]{IEEEtran}
\IEEEoverridecommandlockouts

\usepackage{amsmath,amsfonts}

\usepackage{amssymb}
\usepackage{graphicx}
\usepackage{textcomp}
\usepackage{xcolor}
\usepackage{cite}
\usepackage{xspace}
\usepackage{graphics}
\usepackage{enumitem}
\usepackage{booktabs}
\usepackage{multirow}
\usepackage{bbding}
\usepackage{url}
\usepackage{tikz}
\usepackage{filecontents}
\usepackage[hidelinks]{hyperref}

\usepackage{pifont}

\usepackage{graphicx}
\usepackage{balance}  % for  \balance command ON LAST PAGE  (only there!)
\usepackage{verbatim}
\usepackage{color,xcolor}
\usepackage{array}
\usepackage{bbding}
\usepackage{amsmath}
\usepackage{makecell}
\usepackage{algpseudocode}
\usepackage[linesnumbered,ruled,vlined]{algorithm2e}
\usepackage{caption}
\usepackage{subcaption}

\usepackage{ifthen}
\newboolean{condition}
\setboolean{condition}{true}%将条件设置为 true 或 false

\newcommand{\fzb}[1]{\textcolor{black}{#1}}
\newcommand{\wqg}[1]{\textcolor{black}{#1}}

\newcommand{\Paragraph} [1] {\smallskip\noindent{\bf #1 }}

\newcommand{\system}{EvoRAG\xspace}

\begin{document}

% whether page numbers should be shown or not, use 'plain' for review versions, 'empty' for camera ready
% \pagestyle{empty}
% \clearpage
% \setcounter{page}{1}

% \title{\system: Evolving KG-based RAG with Human Feedback-driven Backpropagation}
% \title{\system: Making Knowledge Graph-based RAG Automatically Evolve through Feedback Backpropagation}
\title{\system: Making KG-based RAG Automatically Evolve through Feedback Backpropagation}

\author{
\IEEEauthorblockN{Zhenbo Fu$^{1}$, Yuanzhe Zhang$^{1}$, Qiange Wang$^{1}$, Hao Yuan$^{1}$, Yuehao Xu$^{1}$, Enze Yi$^{2}$, Yanfeng Zhang$^{1}$, Ge Yu$^{1}$}
\IEEEauthorblockA{\textit{$^{1}$School of Computer Science and Engineering, Northeastern University, Shenyang 110819, China;}\\ \textit{$^2$Northeast Electric Power Research Institute of State Grid Liaoning Electric Power Supply Co, Ltd, Shenyang, 110004, China}}
fuzb@mails.neu.edu.cn, \{zhangyz,yuanhao\}@stumail.neu.edu.cn, thecoldxyh@gmail.com\\
\{wangqg, zhangyf, yuge\}@mail.neu.edu.cn, yienze\_cs@163.com
}

\maketitle

\begin{abstract}

Knowledge Graph-based Retrieval-Augmented Generation (KG-RAG) has emerged as a promising paradigm for enhancing LLM reasoning by retrieving multi-hop paths from KGs. However, existing KG-RAG frameworks often underperform in real-world scenarios because the pre-captured knowledge dependencies are not tailored to the downstream task or its evolving requirements. These frameworks struggle to adapt to task-specific requirements and lack mechanisms to filter low-contribution knowledge during generation. We observe that feedback on generated responses offers effective supervision for improving KG quality, as it provides insights into the correctness and usefulness of the output. However, a key challenge lies in effectively linking response-level feedback to triplet-level contribution evaluation and knowledge updates in the KG. 

In this work, we propose \system, a self-evolving KG-RAG framework that leverages the feedback over generated responses to continuously refine the KG and enhance reasoning accuracy. \system introduces a feedback backpropagation mechanism that attributes feedback to retrieved paths by measuring their utility for response and propagates this utility back to individual triplets, supporting fine-grained KG refinements towards more adaptive and accurate reasoning. Through \system, we establish a closed loop that couples feedback, LLM, and graph data, continuously enhancing the performance and robustness in real-world scenarios. Experimental results show that \system improves reasoning accuracy by $7.34\%$ over state-of-the-art KG-RAG frameworks.

\end{abstract}


\section{Introduction}

Retrieval-Augmented Generation (RAG) \cite{RAGSurvey_arxiv23, RAG_NIPS20, RAGAISurvey_arxiv24} empowers Large Language Models (LLMs) to improve response quality by leveraging external knowledge, and has been widely adopted in various domains \cite{RAGForLegalQA_ICCR24, RAGForHealth_ACL24, RAGForFinacial_ICAIF23}. Among RAG paradigms, Knowledge Graph-based RAG (KG-RAG) \cite{GraphRAGSurvey_arxiv25, GraphRAGsurvey_arxiv_2024, GraphRAGBench_arxiv25} has gained increasing attention for its ability to transform the textual corpus into structured knowledge graphs (KGs) \cite{GraphRAGBench_arxiv25}, capturing rich semantic information and entity-level relations. Given a query $q$, the core idea of KG-RAG is to retrieve a relevant knowledge subgraph (KSG) from the KG and feed it together with $q$ to the LLM for response generation. The retrieved KSG is typically organized into a sequence of reasoning paths, ordered chains of triplets that capture multi-hop semantic connections, where each triplet follows the form $⟨head \ entity, relation, tail \ entity⟩$.

Although KG-RAG frameworks have demonstrated promising results by leveraging the structural information, they often underperform in real-world scenarios due to a mismatch between the pre-captured dependencies in KG and the task-specific requirements \cite{GraphRAGSurvey_arxiv25, GraphRAGBench_arxiv25}. This mismatch primarily stems from two structural limitations of KG-RAG frameworks: insufficient adaptability to downstream reasoning tasks and \fzb{the inability to continuously identify and suppress stale or unreliable knowledge}. These factors are overlooked in recent KG refinement studies~\cite{KGRefine_survey_2016, mlkgr_survey_2024}, which mainly focus on detecting factual errors and semantic inconsistencies. 
Firstly, adaptability refers to the ability to organize and retrieve knowledge in a manner that supports real-time queries. Naively applying RAG with existing KG frameworks often retrieves information that is semantically correct but contributes little to the query. For example, they may retrieve irrelevant triplets or overlook long-range dependencies required for complex reasoning~\cite{RuleRAG_arxiv24, Knowgpt_NIPS24, Gretriever_NIPS24}. This leads to ineffective reasoning and poor knowledge utilization. 
\fzb{Secondly, existing KG-RAG frameworks cannot continuously identify and suppress stale or unreliable knowledge. Since they generally operate over fixed KGs, outdated relations that no longer support downstream applications may retain high retrieval priorities \cite{TCROF_arxiv25, EraRAG_arxiv25}, thereby degrading reasoning performance. These limitations reveal a data management problem: how to continuously refine KGs to improve adaptability and support evolution.}

\begin{figure}
    \centering
    \includegraphics[width=0.9\linewidth]{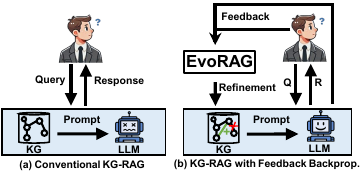}
    \vspace{-0.06in}
    \caption{
    Comparison of conventional KG-RAG and \system.
    \system introduces a backpropagation mechanism that propagates response-level feedback to individual triplets, enabling continuous KG refinements.
    }
    \vspace{-0.2in}
    \label{fig:graphrag_flow}
\end{figure}

\fzb{In real-world deployments, KG-RAG systems typically serve many concurrent users over a shared KG. Queries often concentrate on the same local regions, such as ongoing events and trending topics \cite{yuan2025depcache}. This locality amplifies the aforementioned problems: irrelevant facts and missed long-range dependencies repeatedly affect many users, and outdated triplets continue to mislead multiple responses.}

We observe that feedback derived from generated responses, such as signals on correctness, consistency, and user satisfaction, provides a valuable data source for continuously improving the KG. In LLM-based systems, such feedback is commonly obtained by leveraging LLMs to evaluate generated responses \cite{selfrefine2023nips, reflexion2023nips, critiquellm2024acl, G_eval2023acl, ye2024beyond, trivedi2024self, peng2023check, RLAIFvsRLHF_ICML24}. Similar feedback may also come from human judgments \cite{RLTHF_ICLM, RRHF_NIPS24, KGT_arxiv24, KGV_IPM25, cleangraph_arxiv24} or comparisons with available ground truth \cite{ACE2025arxiv, han2025self, wang2025reinforcement}. Such feedback directly reflects reasoning effectiveness and can serve as a guidance signal for continuously refining the KG to improve task performance.

However, a mismatch in granularity between feedback and KG refinement makes such integration difficult: feedback assesses overall response quality over multiple retrieved paths, whereas KG refinement operates at the level of individual triplets, leading to a mismatch as feedback reflects the combined effect of multiple interacting triplets rather than localized signals for each discrete and reusable knowledge unit. As a result, feedback needs to be correctly attributed to reusable triplets across multiple reasoning paths and queries. Unlike traditional machine learning systems that explicitly construct a computation graph during the forward pass and rely on backpropagation to propagate gradients to model parameters, feedback-driven KG refinement lacks an inherent mechanism to propagate feedback to underlying triplets, requiring carefully designed algorithms to establish this association.

In this work, we present \system, a self-evolving KG-RAG framework that utilizes feedback to continuously \fzb{refine the static KG,} enhancing the accuracy and relevance of the reasoning process, as illustrated in Figure \ref{fig:graphrag_flow}.
To realize this, \system introduces a feedback backpropagation mechanism that establishes a connection between response-level feedback and triplet-level knowledge updates through two key stages. Firstly, \system attributes the feedback to individual reasoning paths by measuring their utility, i.e., how much each path contributes to the response reflected in the feedback. Reasoning paths are used as intermediates because they directly guide the LLM in generating the response, while also providing a natural bridge to the underlying triplets from which they are constructed. Secondly, \system propagates the path utility to update the contribution score of each involved triplet. These scores are maintained as learnable parameters and continuously updated across interactions, allowing \system to prioritize task-relevant knowledge. Through \system, we construct a closed-loop mechanism that tightly couples feedback, LLM reasoning, and triplet-level KG updates, continuously improving its accuracy in real-world queries.

In summary, our primary contributions are as follows: 

\begin{itemize}[leftmargin=*]

    \item \fzb{We propose a feedback backpropagation mechanism that attributes response-level feedback to triplet-level knowledge updates, enabling cross-query feedback reuse.}

    \item \fzb{We develop a feedback-guided KG evolution strategy, which suppresses persistently low-contribution relations and fuses high-contribution multi-hop paths into shortcuts, improving retrieval quality and downstream reasoning accuracy.}

    \item We develop \system, a self-evolving KG-RAG framework that adapts a static KG to evolving task requirements. Experimental results show that \system improves accuracy by $7.34\%$ compared to the SOTA KG-RAG frameworks.
    
\end{itemize}

\section{Preliminary}
\label{sec2}

\subsection{KG-RAG}

\Paragraph{RAG.}
\wqg{Retrieval-Augmented Generation (RAG) \cite{RAGSurvey_arxiv23, RAG_NIPS20, GraphRAGBench_arxiv25, GraphRAGSurvey_arxiv25, neutronrag_demo_sigmodemo25} augments LLMs with external knowledge, enabling responses grounded in up-to-date and task-relevant information. By retrieving relevant evidence from external data, RAG has become a widely adopted paradigm for improving generation quality and reducing hallucinations \cite{llm_hallucination_arxiv23, LLM_hallucination_ACMCS2023, llm_outdated_arxiv24}.}

Among existing RAG methods, KG-RAG \cite{MicrosoftGraphRAG_arxiv_2024, lightrag_arxiv_2024, kag_arxiv24, PathRAG_arxiv25, DALK_EMNLP24, RuleRAG_arxiv24, GNNRAG_arxiv_2024, Graph_CoT_ACL24, Gretriever_NIPS24, Structrag_arxiv24, GraphReader_EMNLP24, StructureGuided_arxiv24, yuan2025depcache} has attracted considerable attention due to its ability to retrieve structured knowledge \cite{wang2022neutronstar, fu2025neutrontask, chen2023neutronstream}. KG-RAG organizes external corpus in a KG, enabling a more comprehensive understanding by leveraging interconnections between knowledge, making it well-suited for complex reasoning tasks.

\Paragraph{KG.}
Knowledge Graph (KG) encodes a wide range of knowledge in the form of triplets: $G = \{(e_1, r, e_2)|e_1, e_2 \in E, r \in R\}$, where $E$ denotes the set of nodes (entities) and $R$ denotes the set of directed edges that signify relations between those entities. The triplet $(e_1, r, e_2)$ is the fundamental building unit of knowledge in a KG, where $e_1$ denotes the subject entity, $r$ denotes the predicate, and $e_2$ denotes the object entity. 
Reasoning path refers to the sequence of entities and relations: $L=e_{0} \xrightarrow{r_{1}} e_{1} \xrightarrow{r_{2}} \cdots \xrightarrow{r_{i}} e_{i}$, where $e_0$ is the starting entity, $e_i \in E$ is the $i$-th entity, and $r_i \in R$ is the $i$-th relation. Such paths reveal higher-order dependencies and provide interpretable evidence for downstream tasks, e.g., link prediction and question answering, and recommendation.

\begin{figure}
    \centering
    \includegraphics[width=0.8\linewidth]{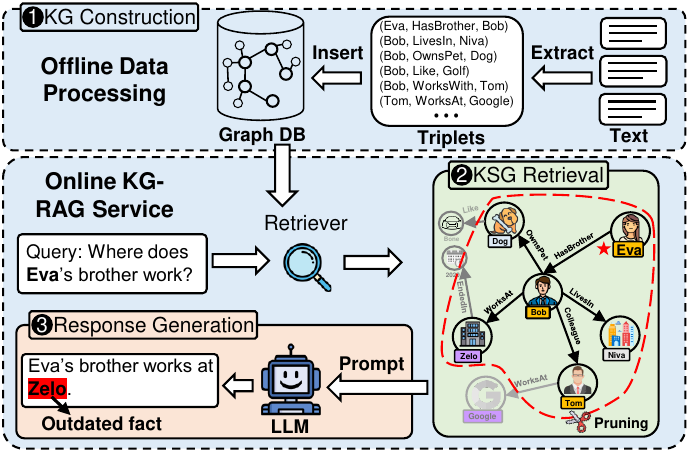}
    \vspace{-0.05in}
    \caption{The overall workflow of KG-RAG.
    }
    \vspace{-0.2in}
    \label{fig:GraphRAG_Workflow}
\end{figure}

\Paragraph{The workflow of KG-RAG.}
\label{sec:2.2}
As illustrated in Figure \ref{fig:GraphRAG_Workflow}, KG-RAG frameworks typically consist of three key components: KG construction, KSG retrieval, and response generation.

(\ding{202}) \textbf{KG construction.} KG construction refers to the process of mining structured knowledge from large-scale texts, which can be performed using specialized information extraction tools, such as OpenIE \cite{OpenIE_ICCL18}, or by leveraging LLMs. Specifically, the text is divided into multiple chunks, each of which undergoes entity recognition and relation extraction to construct triplets, such as $(Tom, WorksAt, Google)$. These triplets are then inserted into a graph database to form a complete KG, which supports subsequent online RAG services.

(\ding{203}) \textbf{KSG retrieval.} KSG retrieval refers to the process of extracting relevant reasoning paths from a KG for a given query, which typically consists of three steps: \texttt{Query Entity Recognition}, \texttt{Subgraph Extraction}, and \texttt{Path Retrieval} \cite{lego-graphrag_vldb25}. \texttt{Query Entity Recognition} operates on the query to identify entities and align them with corresponding nodes in the KG. This step typically employs a semantic model to extract entities whose surface forms or contextual embeddings closely match the query expressions. \texttt{Subgraph Extraction} operates on the entire KG to extract a subgraph related to the query entities. The goal of this step is to reduce the search space and improve the effectiveness and efficiency of retrieval. \texttt{Path Retrieval} operates on the extracted subgraphs to collect the top-$M$ reasoning paths starting from the query entities and connecting them to potential target answers (e.g., one of the reasoning paths for the entity ``$Eva$'' is $Eva \xrightarrow{HasBrother} Bob \xrightarrow{WorksAt} Zelo$). Both \texttt{Subgraph Extraction} and \texttt{Path Retrieval} can be implemented using graph algorithms (e.g., neighbor expansion, personalized PageRank) or semantic models (e.g., embedding models, LLMs). By integrating these approaches, existing KG-RAG frameworks can be broadly covered \cite{kag_arxiv24, PathRAG_arxiv25, DALK_EMNLP24, RuleRAG_arxiv24, GNNRAG_arxiv_2024, Graph_CoT_ACL24, Gretriever_NIPS24, Structrag_arxiv24, GraphReader_EMNLP24, StructureGuided_arxiv24, TOG_ICLR24, Knowgpt_NIPS24, GNNRAG_ISCEIC24}.

(\ding{204}) \textbf{Response generation.} Response generation refers to the process of producing the response conditioned on both the query and the reasoning paths. Specifically, the retrieved paths are combined with the query to form an augmented prompt, which is fed into an LLM for generation. Unlike traditional RAG, which ranks individual text segments by semantic similarity, KG-RAG links entities and relations to compose multi-hop evidence, which enables cross-passage inference and yields more coherent and better grounded responses.

\subsection{Motivation}
\label{sec:motivation}

\begin{figure}
    \centering
    \includegraphics[width=0.8\linewidth]{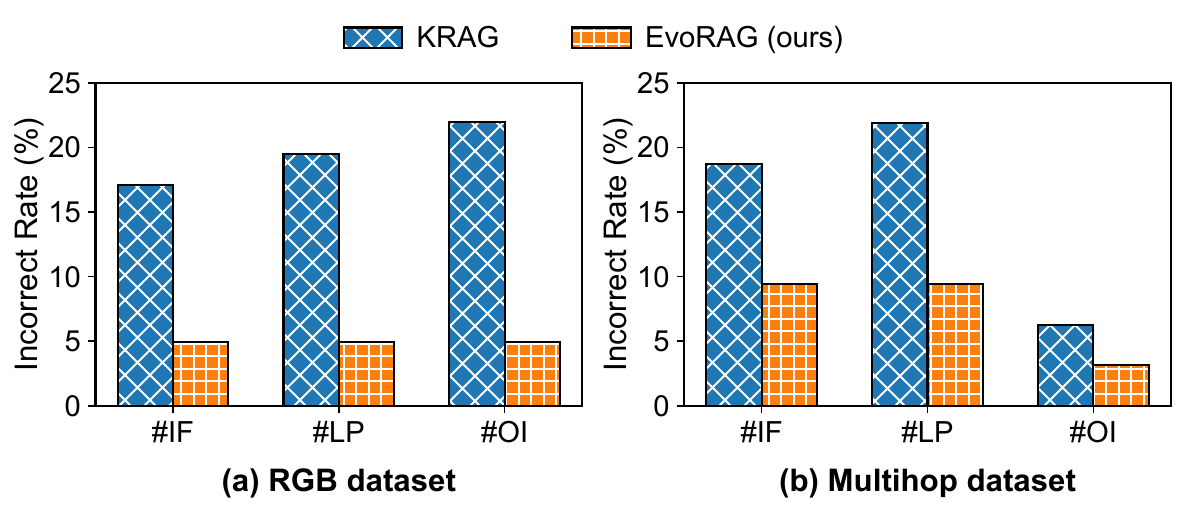}
    \vspace{-0.1in}
    \caption{Proportion of error types in KRAG and \system. \#IF, \#LP, and \#OI represent three error types: irrelevant facts, long reasoning paths, and outdated information, respectively.
    }
    \vspace{-0.15in}
    \label{fig:error_segment}
\end{figure}

KG-RAG has shown great potential in enhancing generation quality by leveraging structural information. However, in real-world applications, a notable gap exists between the KG's predefined dependencies and adaptive, dynamic user requirements. This gap results in two fundamental limitations: lack of adaptability to downstream reasoning tasks and \fzb{inability to identify and suppress unreliable knowledge}. Firstly, the underlying KGs are inherently query-agnostic, as they are constructed offline without considering the specific query intent. This often leads to two common issues: (1) Irrelevant facts introduce noise. As shown in Figure~\ref{fig:GraphRAG_Workflow}, when the user asks “Where does Eva’s brother work?”, the retriever returns unrelated facts such as $(Bob, LivesIn, Niva)$. (2) Failing to capture long-range dependencies. For example, answering that Eva’s brother works at Google requires a $3$-hop reasoning path: $Eva \xrightarrow{HasBrother} Bob \xrightarrow{WorksIn} Team \ \alpha \xrightarrow{PartOf} Google$. However, pre-defined retrieval strategies (e.g., $2$-hop retrieval) fail to capture such multi-hop dependencies, causing critical information to be omitted. \fzb{Secondly, existing KG-RAG systems typically rely on fixed KGs and lack mechanisms to continuously reduce the priority of unreliable triplets} \cite{kag_arxiv24, PathRAG_arxiv25, DALK_EMNLP24, GNNRAG_arxiv_2024, Gretriever_NIPS24}. As a result, outdated or invalid information may persist and mislead the LLM. For example, the KG may still contain $(Bob, WorksAt, Zelo)$, even though the company ceased operations in $2020$.

To systematically quantify the impact of the above limitations, we analyze the erroneous responses in the RGB~\cite{RGB_benchmark_AAAI24} and MultiHop~\cite{Multihoprag_arxiv24} datasets. We categorize the errors into four types: (1) irrelevant facts (IF), (2) long reasoning paths (LP), (3) outdated information (OI), and (4) other errors caused by missing knowledge in the KG or hallucinations by the LLM (not the focus of this work). To provide a baseline, we adopt a state-of-the-art instance from prior work \cite{lego-graphrag_vldb25}, referred to as KRAG. KRAG constructs the retrieval subgraph by expanding from the query entities to include all entities within two hops. Then, a semantic model (BAAI/bge-large-en-v1.5~\cite{bge_arxiv23}) is employed to select reasoning paths within the subgraph that are most similar to the query. As shown in Figure~\ref{fig:error_segment}, IF, LP, and OI together account for over half of all errors, with average proportions of $17.9\%$, $20.7\%$, and $14.1\%$, respectively, making them the primary bottlenecks that limit the effectiveness of KG-RAG. To address these critical limitations, we propose \system, which incorporates a feedback backpropagation mechanism to explicitly target and mitigate these error sources, resulting in substantial improvements in reasoning accuracy. In the following sections, we present the design and implementation of \system in detail.

\section{System Overview}
\label{sec_3_overview}

\begin{figure}
    \centering
    \includegraphics[width=0.8\linewidth]{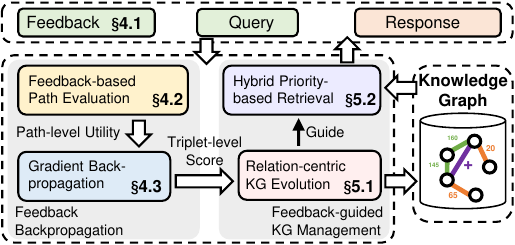}
    \vspace{-0.05in}
    \caption{\system system overview.
    }
    \vspace{-0.2in}
    \label{fig:system_overview}
\end{figure}

We propose \system, a self-evolving KG-RAG framework that continuously improves reasoning effectiveness through fine-grained knowledge refinement and retrieval optimization. As illustrated in Figure~\ref{fig:system_overview}, the system operates in an iterative loop where users continuously submit queries. For each query, \system retrieves multiple reasoning paths from the KG, which are fed into an LLM to generate responses. To further enhance reasoning quality, \system introduces a feedback backpropagation mechanism (see Section~\ref{sec4}) that transforms coarse-grained feedback into fine-grained supervision over individual triplets. This mechanism enables the system to iteratively refine both \fzb{the triplets over a static KG} and the retrieval process. We implement this mechanism in two stages.

First, we propose a feedback-based path evaluation module that propagates feedback to the retrieved reasoning paths. The output is a path-level utility score that quantifies its quality and the contribution to the generated response. Instead of directly evaluating individual triplets, we adopt reasoning paths as the intermediate unit. This design is motivated by the observation that triplets do not function in isolation. During inference, responses are generated through multi-hop reasoning over paths composed of multiple triplets, where the contribution of each triplet is inherently context-dependent, shaped by other triplets within the same path and the semantics of the query. Consequently, directly assigning utility to individual triplets may lead to misalignment, as a triplet that is beneficial in one path can be misleading in another. Reasoning paths therefore provide a more appropriate intermediate abstraction, bridging high-level feedback and fine-grained knowledge units.

Second, we propose gradient backpropagation that propagates path-level utility to the constituent triplets. Specifically, we compute the gradients of the utility expectation over paths with respect to contribution scores and distribute them to triplets based on their roles within each path. To enable consistent supervision across queries, we associate each triplet with a learnable contribution score, which serves as the foundation for fine-grained reasoning supervision and allows the KG to gradually evolve toward task-specific relevance. This score is defined as a dimensionless and non-negative coefficient that estimates the expected influence of a triplet on response quality. In practice, all scores are initialized to $100$ and updated over time based on accumulated feedback signals.

With the updated contribution scores, \system provides feedback-guided KG management, enabling both KG evolution and retrieval optimization. Specifically, \system implements relation-centric KG evolution, which includes relation fusion and suppression. Relation fusion introduces shortcut edges between frequently co-occurring high-utility triplets to enhance long-range reasoning effectiveness, while relation suppression downweights low-utility triplets, ensuring structural coherence and knowledge quality. Then, \system adopts a hybrid priority-based retrieval strategy that integrates semantic relevance with feedback-derived contribution scores. This strategy balances query-dependent relevance and query-independent reliability, guiding retrieval toward both contextually appropriate and historically effective knowledge.

\section{Feedback Backpropagation}
\label{sec4}

\subsection{Feedback}
\label{sec:feedback_generation}

\Paragraph{Overview of feedback and its sources.}
In interactive systems, responses are often accompanied by signals that indicate their quality with respect to task objectives, such as factual correctness, reasoning soundness, or relevance to user intent. We refer to such signals as feedback.
In LLM-based systems, feedback is commonly obtained by using LLMs to evaluate generated responses and produce quality assessments \cite{selfrefine2023nips, reflexion2023nips, critiquellm2024acl, G_eval2023acl, ye2024beyond, trivedi2024self, peng2023check, RLAIFvsRLHF_ICML24}. More generally, similar feedback may also arise from other sources. For example, human assessments may appear as ratings, binary judgments, or survey responses \cite{RLTHF_ICLM, RRHF_NIPS24, KGT_arxiv24, KGV_IPM25, cleangraph_arxiv24}. When ground truth is available, comparison with reference answers provides an objective measure of correctness \cite{ACE2025arxiv, han2025self, wang2025reinforcement}.

In this work, we primarily rely on feedback derived from LLM-based evaluation, following prior studies that use LLMs to assess response quality \cite{selfrefine2023nips, reflexion2023nips, critiquellm2024acl, G_eval2023acl}. Given the ground-truth answers in our datasets, the evaluator LLM is prompted to assess each generated response and assign a scalar score from 1 to 5, providing a reliable and objective assessment. A score of 1 indicates complete dissatisfaction (e.g., irrelevant or incorrect responses), while a score of 5 indicates satisfaction (e.g., responses that are correct, relevant, and well aligned with the query), with intermediate scores ($2$–$4$) reflecting varying levels of partial adequacy. To demonstrate the generality of \system, we further evaluate it using other kinds of feedback, including human judgments and ground-truth based feedback using F1 scores (Section \ref{sec:diff_feedback}).

\Paragraph{Challenge.}
Although unified feedback scoring provides a consistent evaluation signal, a fundamental challenge lies in its granularity mismatch. Feedback is only available at the response level, offering coarse-grained assessments of overall quality, whereas effective KG refinement requires fine-grained identification of the specific triplets that influence reasoning and retrieval. This mismatch limits precise knowledge adjustment and constrains overall system performance.

To address this challenge, our framework translates response-level feedback into fine-grained supervision through a two-step decomposition process. Feedback is first transformed into path-level utility via a path evaluation module, which captures the contribution of entire reasoning trajectories. The utility is then propagated to constituent triplets through gradient backpropagation to update their contribution scores.

\subsection{Feedback-based Path Evaluation}

\label{sec:feedback_backpropagation}

\begin{figure}
    \centering
    \includegraphics[width=0.8\linewidth]{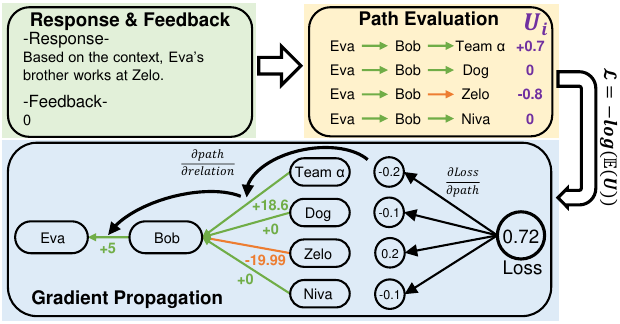}
    \vspace{-0.05in}
    \caption{Feedback Backpropagation. We first use an evaluation function to compute the utility of each reasoning path based on feedback, and then transform it into gradients that are propagated to individual triplets.}
    \vspace{-0.15in}
    \label{fig:feedback_driven_backpropagation}
\end{figure}

To enable effective learning from response-level feedback, we define a utility evaluation function $f$ that assigns utility to each reasoning path by analyzing its contribution to the generated response. The evaluation considers the semantics of the query, the retrieved paths, the response, and the feedback score. Specifically, when feedback indicates a correct response, $f$ rewards paths that support the answer; when the response is incorrect, it attributes negative utility to paths that may have misled generation. Instead of assigning binary judgments, $f$ produces continuous utility in a bounded range (i.e., $[-1, 1]$), enabling nuanced assessment of each path’s contribution. We formalize the path utility evaluation as:

\vspace{-0.1in}
\begin{align}
    U \{L\} = f(q, R_q, L, FS),
    \label{form:utility}
\end{align}

\noindent where $q$ is the query, $R_q$ is the response, $L$ is the set of retrieved paths, and $FS$ is the feedback score. In practice, $f$ can be instantiated by a range of semantic models. In this work, we use an LLM as a constrained path-level scoring function. Given the retrieved paths, the query, the generated response, and the feedback score, the LLM assigns a utility score to each path based on its relevance, following the LLM-as-a-judge paradigm~\cite{TextGrad_nature25, LLM_as_a_judge_NIPS23}. This process is analogous to RAG guided by the external knowledge: the LLM evaluates paths using the provided context instead of producing new responses, and is therefore minimally affected by hallucination.

To capture the multi-faceted nature of path quality under feedback, we decompose path evaluation into three complementary dimensions. \textbf{Supportiveness} measures whether a path provides evidence that supports or undermines the response, conditioned on response correctness as determined by feedback. Higher Supportiveness increases path utility, while lower values decrease it. Fidelity and Conflict serve as auxiliary metrics for cross-validation. \textbf{Fidelity} measures how much a path contributes to the response, with higher values meaning greater contribution. \textbf{Conflict} measures whether a path contradicts the response, with lower values meaning less contradiction. 
\fzb{The LLM assigns a Supportiveness score $S\in[-1,1]$ and two binary gates, Fidelity(F) and Conflict(C), in $\{-1,1\}$. Importantly, a path’s utility is updated based on Supportiveness only when F=1 and C=-1; otherwise, the update is suppressed.}
This design prevents erroneous utility updates caused by spurious grounding or contradictory evidence, ensuring reliable path-level attribution.

\subsection{Gradient Backpropagation}

Once the utility of each reasoning path is evaluated, the next step is to propagate this utility back to individual triplets, refining their contribution scores so that future retrievals are more likely to select high-utility paths. To derive backpropagation, we first need to introduce how the contribution score influences the path selection strategy during forward computation.

\subsubsection{Forward Computation} 
In the forward retrieval phase, each query retrieves multiple reasoning paths from the KG, which is composed of a sequence of triplets. Each triplet $t$ is associated with a selection probability ($P(t) \in [0, 1]$), which reflects its semantic relevance to the query and the contribution score. Formally, the $P(t)$ can be defined as:
\begin{align}
P(t)=(1 - \alpha) S_r(t)+\alpha S_c(t),    
\label{formula:P(t)}
\end{align}

\noindent where $S_r(t) \in (0, 1]$ measures the semantic similarity between the triplet $t$ and the query $q$, providing a query-specific score that aligns path selection with the query. In contrast, $S_c(t) \in [0, 1]$ is the learnable contribution score by normalization, which accumulates feedback across queries, thereby decoupling triplet assessment from any single query. The parameter $\alpha \in [0, 1]$ is a learnable trade-off between $S_r$ and $S_c$, allowing the system to balance the current query’s guidance with accumulated knowledge from previous feedback.

Based on these probabilities, the priority of a reasoning path ($L_i \in L$) can be defined as follows:

\begin{equation}
\fzb{P(L_i) =\frac{\exp\Big(\tfrac{1}{|L_i|}\sum_{t \in L_i} \log \left(\hat{P}(t)\right)\Big)}{\sum_{L_j \in L}\exp\Big(\tfrac{1}{|L_j|}\sum_{t \in L_j} \log \left(\hat{P}(t)\right)\Big)},
\label{formula:P(L_i)}
}
\end{equation}

\noindent where $L_i$ denotes a path, $|L_i|$ is its hop length (i.e., the number of triplets it contains). \fzb{$\hat{P}(t) = max\{P(t), \epsilon\}$, where $\epsilon$ is a small positive constant used for numerical stability.} The length normalization takes the log-average of triplet probabilities to eliminate the inherent bias of the multiplicative formulation toward shorter paths, enabling fair comparison across paths of different hop lengths.

The Formula \ref{formula:P(L_i)} not only ensures relevance to the current query, but also enables generalization to unseen queries. When a new query arrives, the similarity score $S_r$ is freshly calculated, ensuring specificity to the user’s intent and filtering out irrelevant knowledge, where the contribution score $S_c$ transfers the accumulated feedback on triplet quality to guide the selection of reasoning paths. In this way, the model adapts to unseen queries by letting $S_r$ provide query-time specificity and $S_c$ provide cross-query transferability, yielding path priorities that balance immediate relevance with long-term utility.

\Paragraph{Loss function.}
We formulate an objective that maximizes the expected utility of the retrieved reasoning paths:

\begin{align} 
    \fzb{\mathcal{L} = - \log(max\{{A}, \epsilon\}),\ \ \
    A = \sum_{L_i\in L} P(L_i) \frac{U(L_i)+1}{2},}
    \label{form:J}
\end{align}

\noindent where $L$ denotes the set of all paths. \fzb{$(U(L)+1)/2$ maps utility from [-1,1] to [0,1]. The logarithmic loss maximizes expected normalized utility and provides stronger corrective gradients when the expectation is low.}

\subsubsection{Backward Computation}
\label{backward}
We compute the gradient of the loss $\mathcal{L}$ with respect to the contribution score $S_c(t)$ as follows:

\small
\begin{align}
    \nabla_{S_c(\tau)}\mathcal{L}&=\begin{cases}-\dfrac{\alpha}{2A\cdot P(\tau)}\displaystyle\sum_{i:\tau\in L_i}V_i,& A>\epsilon\ \text{and}\ P(\tau)>\epsilon,\\[10pt]0,& A\leq\epsilon\ \text{or}\ P(\tau)\leq\epsilon.\end{cases}\\
    V_i &= \frac{P(L_i)}{|L_i|}\left(U(L_i) - \sum_{L_j\in L}P(L_j)U(L_j)\right),
\end{align} \normalsize

\noindent where $V_i$ denotes the deviation of a path's utility from the expected utility across all paths. The gradient $\nabla_{S_c(t)} \mathcal{L}$ quantifies how the triplet $t$ contributes to reasoning paths with above- or below-average utility. If $t$ frequently appears in high-utility paths (i.e., paths where $U(L_i) > \sum_{L_j\in L}P(L_j)U(L_j)$), the gradient becomes negative, reducing the loss and thereby increasing the contribution score $S_c(t)$. Conversely, if $t$ tends to appear in low-utility paths, the score will be reduced. Similarly, the gradient of $\alpha$ can be computed as follows:

\vspace{-0.1in}
\small
\begin{align}
   \fzb{\nabla_{\alpha}\mathcal{L}=\frac{1}{2A}\sum_{L_i\in L}V_i\sum_{\tau\in L_i,\;P(\tau)>\epsilon}\frac{S_r(\tau)-S_c(\tau)}{P(\tau)}.}
   \label{formula:gradient_alpha}
\end{align}\normalsize

\noindent \fzb{This gradient applies when $A>\epsilon$; otherwise, it is 0. Within each path, terms with $P(\tau)\leq\epsilon$ contribute 0 because of the numerical floor.} Finally, we apply gradient descent to update the contribution score and parameter $\alpha$:

\vspace{-0.15in}
\small
\begin{align}
    S_c(t)&=S_c(t)-\eta\nabla_{S_c(t)}\mathcal{L},\\
   \alpha&=\alpha-\eta\nabla_{\alpha}\mathcal{L},
\end{align}\normalsize

\noindent where $\eta$ is the learning rate. \fzb{$S_c(t)$ is normalized from the raw contribution score and clipped after each update, preserving its valid score range and bounding the influence of extreme updates. The parameter $\alpha$ is clipped to $[0,1]$.} This optimization process ensures that triplets associated with helpful reasoning paths become more prominent, while those consistently tied to unproductive reasoning are gradually suppressed.

\fzb{
These updates occur continually during task execution rather than through one-shot global KG optimization. In regions receiving frequent feedback, \system accumulates triplet contributions to adapt retrieval; in unseen regions, it initially relies on semantic relevance and adapts as feedback accumulates.
}

\subsubsection{Accuracy Analysis}
To evaluate the effectiveness of our feedback backpropagation, we analyze both its convergence and its behavior under noisy feedback. The convergence analysis provides theoretical guarantees for stable and correct updates, while the noise tolerance analysis considers the impact of occasional erroneous feedback on the updates.

\Paragraph{Convergence analysis.}
\fzb{Since $\big(U(L)+1\big)/2$ is in the range of $[0,1]$ and $\sum_{L_i\in L}P(L_i)=1$, the expected normalized utility is also bounded within $[0,1]$. Together with the positive lower bound used in the loss, this ensures that $\mathcal{L}$ is finite and bounded, providing a stable optimization objective. The feasible domain of $S_c(t)$ and $\alpha$ is compact and convex \cite{boyd2004convex}. When the numerical floors are inactive, the loss is smooth with Lipschitz-continuous gradients over the bounded parameter domain \cite{nesterov2013introductory}. Clipping keeps the iterates within this compact feasible domain. Under standard bounded-variance and step-size conditions, the resulting clipped stochastic updates converge to a stationary point \cite{bottou2018optimization}. The resulting updates progressively promote triplets associated with higher-utility paths and suppress those associated with lower-utility paths.}

\Paragraph{Noise tolerance.}
Feedback backpropagation is inherently noise-tolerant to erroneous feedback and LLM scoring errors in two ways. First, contribution scores are updated cumulatively across queries, allowing occasional errors to be averaged over time. 
Second, path utility is updated by supportiveness (the feedback-driven dimension) only when fidelity is 1 and conflict is -1; otherwise, the update is suppressed.
This cross-validation mechanism prevents spurious feedback or LLM misjudgments from distorting path evaluation, while reinforcing consistently supported paths.

\subsubsection{Complexity Analysis}
We analyze the time and space complexity of \system by considering two components. For feedback backpropagation, contribution scores are propagated along each reasoning path and aggregated to update triplets. \fzb{The computation scales with the number of paths $P$, the average path length $H$, and the number of triplets $T$, yielding a time complexity of $\mathcal{O}(PH + P + T)$.} The memory is dominated by storing path utilities and triplet-level contribution scores, yielding a space complexity of $\mathcal{O}(P + T)$. For path evaluation, all reasoning paths are processed in a single LLM call. Let $C_{\text{LLM}}(L)$ denote a black-box cost function of an LLM call with input length $L$, abstracting away model architecture and deployment details. Since $L$ is proportional to the length of all paths, i.e., $L = \mathcal{O}(HP)$, the overall time complexity of path evaluation is $\mathcal{O}(C_{\text{LLM}}(HP))$, while the space overhead is dominated by the prompt and intermediate LLM states.

\section{Feedback-guided KG Management}
\label{sec:contribution_score}

\subsection{Relation-centric KG Evolution over a Static KG}

We refine the KG based on feedback-derived contribution scores that reflect how each triplet supports reasoning. Instead of updating entities, \system adopts a relation-centric evolution strategy. Since entities are shared across multiple triplets, modifying them would introduce cascading and ambiguous effects. In contrast, each relation uniquely defines the semantic intent of a triplet and determines whether the connection between two entities is meaningful, which aligns precisely with what the contribution score measures.

The KG evolution is performed after each feedback iteration, which consists of multiple batches of user queries. After aggregating triplet-level scores, we calculate the global mean $\mu$ and standard deviation $\sigma$, which define refinement thresholds: triplets with scores greater than $\tau_{high} =  \mu + \sigma$ are considered high quality, while those consistently below $\tau_{low} = \mu - \sigma$ throughout iterations are considered low quality. These thresholds provide a data-driven basis, ensuring that relation fusion and pruning are guided by statistically significant differences rather than random fluctuations.

\begin{figure}
    \centering
    \includegraphics[width=0.8\linewidth]{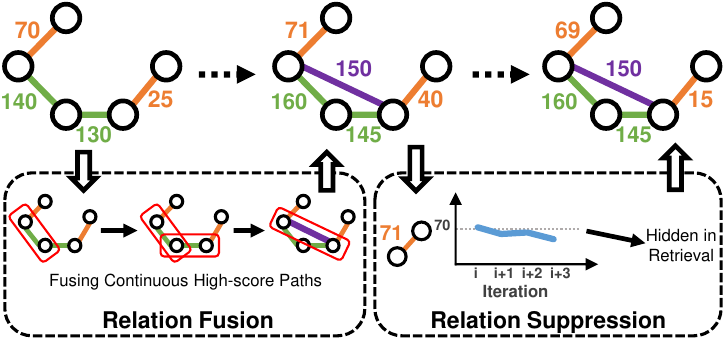}
    \vspace{-0.05in}
    \caption{Illustration of relation-centric KG evolution, where triplets with low contribution scores are progressively down-weighted in retrieval probability.
    \vspace{-0.2in}
    }
    \label{fig:KG_evolution}
\end{figure}

Guided by these thresholds, the KG is evolved through two operations: \textbf{Relation Fusion} and \textbf{Relation Suppression}. As shown in Figure \ref{fig:KG_evolution}, high-quality relations are reinforced by adding shortcuts that connect the endpoints of paths, while persistently low-quality relations are suppressed via reduced contribution scores. Together, these operations balance adaptivity with stability, ensuring that the KG evolves in alignment with feedback while maintaining structural coherence.

\begin{itemize}[leftmargin=*]
\item \textbf{Relation fusion.} This operation strengthens high-quality relations by abstracting a shortcut that connects the endpoints of a multi-hop path, effectively reducing reasoning depth. For a path $L_i = (t_1, t_2, \dots, t_k)$ whose triplets satisfy $S_c(t_i) > \tau_{high}$, \system abstracts a shortcut edge $\hat{r}$:

\vspace{-0.15in}
\begin{align}
(e_0,r_1,e_1), \dots, (e_{k-1}, r_k, e_k) \Rightarrow (e_0, \hat{r}, e_k),
\end{align}

\noindent where $\hat{r}$ is assigned a label and a score: the label is recommended by the LLM based on the semantics of the path, and the score is set to the path’s average contribution.
\fzb{
LLM-generated labels may introduce hallucinated relations. However, feedback backpropagation provides a self-correction mechanism: harmful shortcuts receive lower contribution scores, reducing their retrieval priority, and persistently harmful shortcuts are progressively suppressed. Thus, shortcut influence is filtered by downstream effects rather than reinforced unconditionally.
}

\item \Paragraph{Relation Suppression.} 
We progressively suppress low-quality triplets based on their long-term contribution patterns. Low-contribution triplets are softly deprioritized during retrieval. As defined in Formula \ref{formula:P(t)}, retrieval is jointly guided by semantic similarity and contribution score, which reduces the retrieval probability of low-score triplets without immediately discarding them. Triplets that are semantically relevant to other queries may still be retrieved and can regain contribution once they prove useful again (Section \ref{sec:5.2}).

\end{itemize}

Algorithm~\ref{alg:KGR} summarizes the KG evolution process. Given the KG $\mathcal{G}^{(h)}$, we first compute the mean $\mu$ and standard deviation $\sigma$ of triplet contribution scores $S_c^{(h)}(t)$ over all triplets $t \in \mathcal{G}^{(h)}$ (line 1). Then, we start from triplets whose scores exceed the threshold (line 3), and perform parallel BFS to identify candidate multi-hop paths (lines 4-16). \fzb{A shortcut edge is created between the endpoints of a path if (i) every triplet contribution score in the path exceeds $\tau_{\mathrm{high}}$ and (ii) no existing edge connects the endpoints in the KG (lines 11-15).} To improve efficiency, neighbors are traversed in descending order of $S_c^{(h)}$ (line 9), allowing early termination within each BFS layer when remaining neighbors fall below the threshold (line 12). This process ensures that the KG gradually refines its structure, strengthening consistently useful relations while suppressing noisy or unreliable ones.

\fzb{It is worth noting that \system performs feedback-guided relation evolution over a static base KG, rather than dynamic KG maintenance. This relation-centric design reflects the supervision available in our setting: feedback reveals how existing relations contribute to downstream reasoning, but does not provide sufficient evidence to verify entity changes, newly arriving facts, or factual corrections. Accordingly, contribution scores adjust the structural importance of existing relations to improve reasoning behavior. Future work will extend feedback backpropagation to dynamic KGs by incorporating externally verified entity and fact updates.}

\begin{algorithm}[t] \small
    \caption{Feedback-driven KG Evolution.}
    \label{alg:KGR}
    \KwIn{KG $\mathcal{G}^{(h)}$ at iteration $h$, contribution scores $S_c^{(h)}$ for each triplet $t$ appearing in $\mathcal{G}$, max hop $H$}
    \KwOut{Updated KG $\mathcal{G}^{(h+1)}$}
     Compute global mean $\mu$ and std $\sigma$ of $S_c^{(h)}$;\\
     Initialize $Shortcut = []$;\\
     $T_{start} \gets \{t | t \in \mathcal{G}$ \textbf{and} $S_c^{(h)}(t) > \mu + \sigma\}$;\\  
    \textbf{parallel} \For{$t_{start} \in T_{start}$}{
    $Frontier \gets [t_{start}]$; \\
        \For{$hop = 1 \to H$ \textbf{and} $Frontier \neq []$}{
            $NextFrontier \gets []$; \\
            \For{each $t_{curr} \in Frontier$}{
                $N \gets \textsc{NbrTriplet}(t_{curr})$; \textcolor{gray}{// sorted by $S_c^{(h)}$ to prioritize high-contribution relations} \\
                \For{each $t_{nbr} \in N$}{
                    \fzb{\If{$S_c^{(h)}(t_{\mathrm{nbr}})\leq\tau_{\mathrm{high}}$}{
                        \textbf{break};
                    }
                    \If{$\exists(t_{\mathrm{start}}.\mathrm{head},r,
                    t_{\mathrm{nbr}}.\mathrm{tail})\in KG$}{
                        \textbf{continue};
                    }
                    $\bar{S}\leftarrow
                    \operatorname{Score}(t_{\mathrm{start}},t_{\mathrm{nbr}})$;
                    $\mathrm{Shortcut.Push}
                    (((t_{\mathrm{start}}.\mathrm{head},r^*,
                    t_{\mathrm{nbr}}.\mathrm{tail}),\bar{S}))$;
                    $\mathrm{NextFrontier.Push}(t_{\mathrm{nbr}})$;}
                }
            }
            $Frontier \gets NextFrontier$; \\
        }
    }

    $G^{(h+1)} \gets (G^{(h)} \cup Shortcut)$;\\
    \Return $G^{(h+1)}$

\end{algorithm}

\subsection{Hybrid Priority-based Retrieval}
\label{sec:5.2}

\system introduces a hybrid priority-based retrieval module that jointly leverages semantic relevance and feedback-derived triplet contribution. Building on existing KG-RAG retrieval methods \cite{lego-graphrag_vldb25, lightrag_arxiv_2024, MicrosoftGraphRAG_arxiv_2024} that emphasize short-term semantic relevance, our approach incorporates long-term feedback to progressively adjust retrieval priorities.

When a query $q$ arrives, \system first encodes it into an embedding vector and retrieves the top-$N$ relevant entities ($E_q$) as starting nodes based on cosine similarity. The $k$-hop neighborhoods of these entities are traversed to construct a high-recall subgraph that preserves useful multi-hop relations. This step ensures comprehensive coverage of potentially useful knowledge.
Within the extracted subgraph, \system performs a hybrid ranking of candidate reasoning paths. Each triplet $t$ is associated with two complementary scores: (1) a relevance score $S_r(t)$ that measures semantic similarity to $q$, and (2) a contribution score $S_c(t)$ that captures its historical utility across iterations. These scores are integrated into a unified path priority $P(L_i)$ (Eq.~\ref{formula:P(L_i)}), which adaptively balances short-term relevance with long-term reliability. For each starting entity $e \in E_q$, \system retains the top-$M$ paths with the priorities, thereby suppressing those dominated by low-quality triplets and emphasizing reusable, high-confidence knowledge.

Importantly, \system remains compatible with various retrieval backends. While $S_r(t)$ currently follows a similarity-based formulation \cite{lego-graphrag_vldb25, lightrag_arxiv_2024, MicrosoftGraphRAG_arxiv_2024}, it can be replaced with other strategies such as ToG \cite{TOG_ICLR24} or DALK \cite{DALK_EMNLP24}. The key innovation lies in the integration of $S_c(t)$, which injects feedback-driven triplet quality into the retrieval process, forming a flexible and adaptive hybrid retrieval paradigm.

Empirically, setting $N = 10$ and $M = 10$ yields the best performance, achieving high response accuracy. Larger values bring marginal gains while increasing retrieval cost (Section \ref{sec:secsitive_study}). This hybrid retrieval improves reasoning accuracy by filtering low-contribution triplets and reducing unnecessary context, resulting in more efficient and focused LLM input.

\section{EXPERIMENTAL EVALUATION}

\subsection{Experimental Setup}
\label{sec:4.1}

\Paragraph{Environments.}
The experiment is conducted on a GPU server equipped with $2$ Intel(R) Xeon(R) Silver $4316$ CPUs, $503$GB DRAM, and $2\times$ NVIDIA RTX A6000 ($48$ GB) GPUs.

\Paragraph{Datasets.}
We evaluate \system on three real-world QA datasets summarized in Table~\ref{tab:Dataset}: RGB~\cite{RGB_benchmark_AAAI24}, MultiHop (MTH)~\cite{Multihoprag_arxiv24}, and HotpotQA (HPQ)~\cite{Hotpotqa_EMNLP18}. We use all $300$ English-language reasoning queries from RGB and all $816$ reasoning queries from MTH. Following prior work~\cite{TCROF_arxiv25}, we randomly sample $600$ HPQ queries to limit evaluation cost.

\Paragraph{Training and test set construction.}
To evaluate \system under realistic online-service conditions, we simulate a multi-user scenario where numerous queries concentrate on the same local regions of the KG. Based on this setting, we construct the training and test sets as follows. The test set directly uses the original queries listed in Table~\ref{tab:Dataset}, representing user queries in these hotspot regions. The training set is constructed by generating additional query–answer pairs within the subgraphs retrieved based on the test queries. For each subgraph, we sample alternative reasoning paths (up to two hops) that differ from the paths used to form test queries, avoiding data leakage. The corresponding target entities are used as ground-truth answers, producing diverse yet localized queries that simulate multi-user access to the same hotspot regions. During training, the generated queries are issued over multiple iterations to simulate real-time user interactions, allowing the system to collect feedback and update the KG incrementally. The model’s performance is then evaluated on the test set.

\begin{table}
\small
\caption{Dataset description and the corresponding sizes of constructed KGs.}
\label{tab:Dataset}
\centering
{\renewcommand{\arraystretch}{1.2}
\footnotesize
\begin{tabular}{c|c|c|c|c}
\hline

\hline
{\textbf{Dataset}} & {\textbf{Query}} & {\textbf{Type}} & {\textbf{Entity}}  & {\textbf{Triplets}} \\
\hline

\hline
RGB \cite{RGB_benchmark_AAAI24} & 300 & Single-hop & 54,544 & 74,394\\

Multihop (MTH) \cite{Multihoprag_arxiv24} & 816 & Multi-hop & 30,953 & 26,876\\

HotpotQA (HPQ) \cite{Hotpotqa_EMNLP18} & 600 & Multi-hop & 76,280 & 74,942\\

\hline

\hline
\end{tabular}
}

\end{table}

\begin{table}
    \centering
    \caption{Alignment ratio between generated feedback and ground truth.}
    \footnotesize
    \vspace{-0.05in} 
    \begin{tabular}{c|c|c|c}
        \hline
        
        \hline
        \textbf{Dataset} & \textbf{Total Feedback} & \textbf{Correct Feedback} & \textbf{Proportion} \\
        \hline
        \textbf{RGB} & 300 & 288 & 96\% \\
        \textbf{MTH} & 816 & 783 & 95.96\% \\
        \textbf{HPQ} & 600 & 529 & 88.17\% \\
         \hline

         \hline
    \end{tabular}
    \vspace{-0.15in}
    \label{tab:Statistics_of_simulated_feedback}
\end{table}

\begin{table*}[t]
    \centering
    \caption{
    Performance comparison with KG-RAG and KGR methods. TransE, RotatE, CAGED, and LLM\_Sim are KGR methods applied to refine the KG, whose effects are evaluated using KRAG due to its operation at the triplet-level retrieval granularity. Best results are highlighted in bold. ↑ indicates the improvement over baselines.
    }
    \vspace{-0.06in}
    \begin{tabular}{c|c|c|c|c|c|c|c|c|c}
        \hline
        
        \hline
        \multirow{2}{*}{\textbf{Dataset}} & \multirow{2}{*}{\textbf{Metric}} &
        \multicolumn{3}{c|}{\textbf{Existing KG-RAG frameworks}} &
        \multicolumn{4}{c|}{\textbf{KRAG with various KGR methods}} &
        \multirow{2}{*}{\textbf{\system}} \\
        \cline{3-9}
         & & \textbf{MRAG} & \textbf{LRAG} & \textbf{KRAG} & \textbf{TransE} & \textbf{RotatE} & \textbf{CAGED} & \textbf{LLM\_Sim} & \textbf{} \\
        \hline
        
        \hline
        \multirow{3}{*}{\textbf{RGB}} & \#ACC & $75.67\%$ & $76.00\%$ & $71.00\%$ & $67.67\%$ & $74.33\%$ & $73.33\%$ & $67.33\%$ & \textbf{84.00\%} {$\uparrow$8.00-16.67} \\
        & \#EM & $47.33\%$ & $47.00\%$ & $42.33\%$ & $42.67\%$ & $47.33\%$ & $45.67\%$ & $43.33\%$ & \textbf{56.67\%} {\small$\uparrow$9.34-14.34} \\
        & \#F1 & $68.69\%$ & $64.99\%$ & $64.21\%$ & $60.61\%$ & $68.47\%$ & $67.02\%$ & $62.95\%$ & \textbf{75.40\%} {\small$\uparrow$6.71-14.79} \\
        \hline
        \multirow{3}{*}{\textbf{MTH}} & \#ACC & $75.61\%$ & $76.20\%$ & $74.02\%$ & $50.12\%$ & $76.47\%$ & $76.72\%$ & $72.06\%$ & \textbf{80.26\%} {\small$\uparrow$3.54-30.14} \\
        & \#EM & $69.61\%$ & $70.43\%$ & $71.94\%$ & $50.37\%$ & $72.55\%$ & $74.02\%$ & $71.45\%$ & \textbf{78.55\%} {\small$\uparrow$4.53-28.18} \\
        & \#F1 & $75.86\%$ & $74.87\%$ & $74.87\%$ & $52.23\%$ & $77.08\%$ & $77.83\%$ & $73.27\%$ & \textbf{80.80\%} {\small$\uparrow$2.97-28.57} \\
        \hline
        \multirow{3}{*}{\textbf{HPQ}} & \#ACC & $38.83\%$ & $44.83\%$ & $39.00\%$ & $27.67\%$ & $36.33\%$ & $29.17\%$ & $32.83\%$ & \textbf{48.16\%} {\small$\uparrow$3.33-20.49} \\
        & \#EM & $25.50\%$ & $25.67\%$ & $24.83\%$ & $17.83\%$ & $24.50\%$ & $20.17\%$ & $21.50\%$ & \textbf{35.84\%} {\small$\uparrow$10.17-18.01} \\
        & \#F1 & $37.36\%$ & $40.47\%$ & $41.32\%$ & $29.34\%$ & $38.51\%$ & $32.21\%$ & $36.07\%$ & \textbf{46.55\%} {\small$\uparrow$5.23-17.21} \\
        \hline
        
        \hline
    \end{tabular}
    \vspace{-0.2in}
    \label{tab:unified_kgr_rag_comparison}
\end{table*}

\Paragraph{Feedback generation.}
By default, we use LLM-generated feedback produced by Qwen2.5-32B (Section~\ref{sec:feedback_generation}) in experiments. Table~\ref{tab:Statistics_of_simulated_feedback} shows that the generated feedback achieves an agreement ratio of $93.38\%$ with the ground truth, indicating that it serves as a reliable proxy for evaluation. To demonstrate the robustness and applicability of \system, we also evaluate it under diverse feedback settings. We first introduce partially noisy feedback to assess its tolerance to unreliable signals (Section \ref{sec:tolerance}). We then consider alternative feedback sources, including human judgments to reflect practical scenarios without explicit ground truth, and ground-truth-based feedback (e.g., F1 score) as an ideal reference (Section \ref{sec:diff_feedback}). The results show that \system maintains stable effectiveness under various feedback reliability and availability settings.

\Paragraph{Baseline methods:}

\begin{itemize}[leftmargin=*]

\item \textbf{KG-RAG framework.} We compare \system with three representative KG-RAG baselines, including Microsoft GraphRAG (MRAG) \cite{MicrosoftGraphRAG_arxiv_2024}, LightRAG (LRAG) \cite{lightrag_arxiv_2024}, and KRAG. MRAG and LRAG are two variants that integrate textual chunks with KGs, providing fine-grained knowledge. MRAG applies community detection algorithms to summarize subgraph information, providing abstract representations. LRAG adopts a dual-level retrieval to combine entity-level and relation-level information. In addition, we adopt KRAG, the most effective KG-RAG instance identified in Lego-GraphRAG \cite{lego-graphrag_vldb25} (see Section \ref{sec:motivation}). 

\item \textbf{KG-RAG with KGR methods.} 
We compare \system with KRAG enhanced by KG refinement (KGR) methods. We integrate several representative KGR methods into KRAG, including deep learning-based methods such as TransE~\cite{TransE_NIPS13}, RotatE~\cite{RotatR_ICLR19}, and CAGED~\cite{CAGED_CIKM22}, and an LLM-based method, LLM\_Sim~\cite{LLMKGR_GenAIK25}. These methods aim to identify and correct noisy or irrelevant triplets.
    
\end{itemize}

\Paragraph{KG construction.}
We construct KGs following prior work \cite{MicrosoftGraphRAG_arxiv_2024, lightrag_arxiv_2024,yuan2025depcache}. Texts are split into 512-token chunks, from which GPT-4o-mini extracts entities and relations using predefined prompts. 
Table~\ref{tab:Dataset} reports KG statistics.
MRAG and LRAG further build higher-level textual representations, while KRAG and \system operate directly on triplet-based KGs.

\Paragraph{Implementation details.}
For each query, KRAG extracts $N=10$ entities and collects their $2$-hop neighbors to form a candidate subgraph. From this subgraph, up to $M=10$ reasoning paths per entity are selected. \system builds upon KRAG by introducing a feedback backpropagation mechanism that learns a contribution score for each triplet (with a learning rate of $0.5$) during online operation. For KGR methods, TransE and RotatE identify noisy triplets using the constraint $||(h + r - t)|| < \gamma$, where $h$ and $t$ denote the embeddings of head and tail entities, and $r$ is the relation embedding. We set $\gamma = 0.1$ following \cite{LLMKGR_GenAIK25}. To ensure fair comparison, we adopt Qwen2.5-32B as LLM backbone across all methods. It is used for path evaluation and response generation.

\subsection{Overall Comparison}

We compare \system with various KG-RAG frameworks and KRAG enhanced by different KGR methods, using accuracy (ACC), exact match (EM), and F1 as metrics. ACC measures the proportion of queries for which the response contains the ground-truth. EM indicates whether the response exactly matches the ground-truth. F1 evaluates the token-level overlap by computing the harmonic mean of precision and recall. \system is trained on the training set, enabling it to adaptively refine the KG. Table \ref{tab:unified_kgr_rag_comparison} reports the results.

Compared to KG-RAG frameworks, \system achieves an average improvement of $7.34\%$ in ACC, $9.84\%$ in EM, and $7.29\%$ in F1 score across three datasets. 
\fzb{These gains stem from adapting retrieval and evolving a static KG through relation suppression and shortcut augmentation.}
Existing KG-RAG frameworks rely on static KGs, where irrelevant or noisy knowledge may suppress useful reasoning paths, limiting performance. In contrast, \system leverages a feedback backpropagation mechanism that propagates response-level feedback to triplets, addressing cases where baselines miss correct answers. In addition, the KG evolution mechanism enhances the KG’s adaptability to RAG tasks, allowing it to evolve continuously during online service.

\begin{figure}
    \centering
    \includegraphics[width=0.8\linewidth]{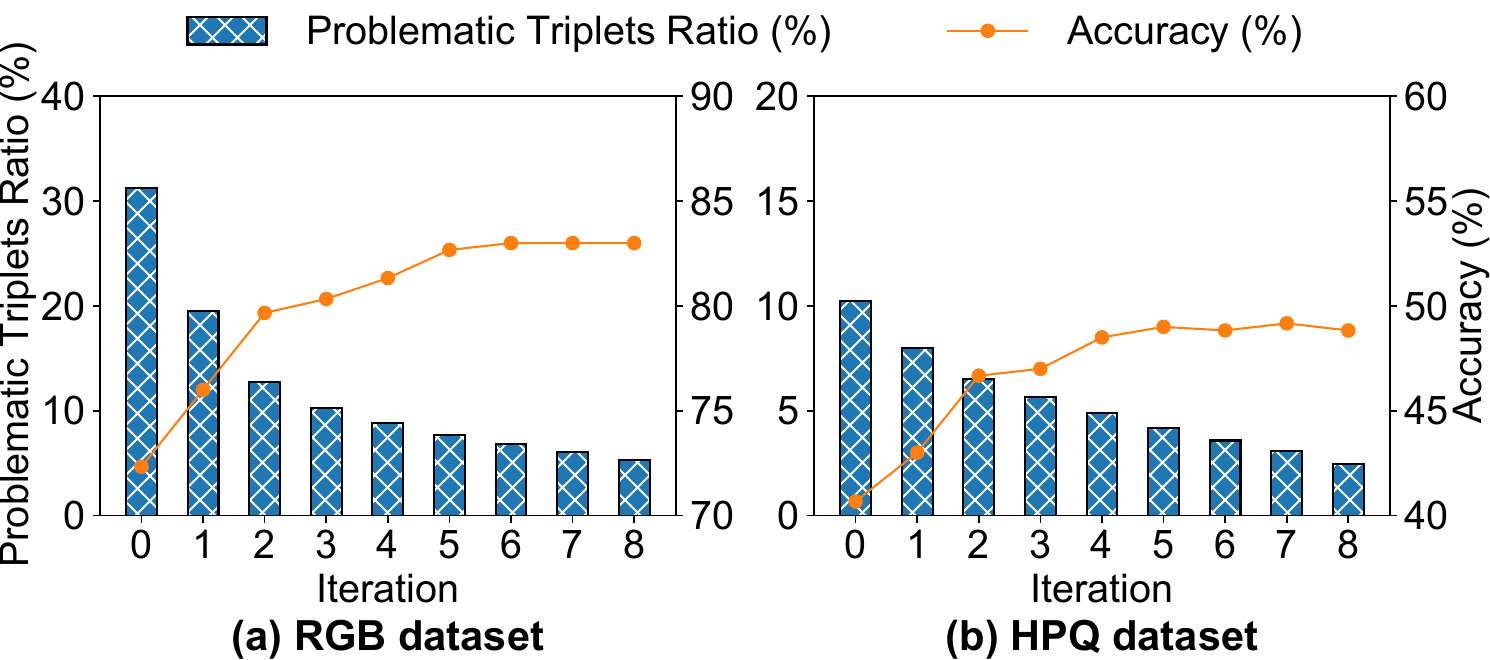}
    \vspace{-0.08in}
    \caption{The problematic triplets ratio and response accuracy across iterations.}
    \vspace{-0.2in}
    \label{fig:noise_degrad}
\end{figure}

Compared to KGR-based KGRAG methods, \system achieves an average improvement of $13.80\%$ in ACC, $12.74\%$ in EM, and $11.28\%$ in F1 score across three datasets. DL-based KGR methods, such as TransE, RotatE, and CAGED, rely on supervised learning to detect noisy triplets. These methods require fine-grained labels indicating the correctness of individual triplets, which are typically unavailable in KG-RAG scenarios. LLM\_Sim attempts to overcome this limitation by leveraging LLMs to evaluate and refine triplets based on internal knowledge. However, it suffers from hallucinations, especially when encountering knowledge beyond the model's training data, and may introduce new noise or mistakenly modify correct facts. 
\wqg{In contrast, \system directly leverages real task feedback as the supervision signal, rather than relying on manually labeled triplets or LLM-generated correctness judgments. By propagating query-level feedback back to the supporting knowledge, \system continuously refines the knowledge graph according to actual reasoning outcomes. This makes the refinement process both more practical and more reliable, avoiding the annotation cost of supervised methods and the hallucination risk of LLM-based approaches.}

\begin{figure*}
    \centering
    \begin{minipage}[t]{0.24\linewidth}
        \centering
        \includegraphics[width=0.95\linewidth]{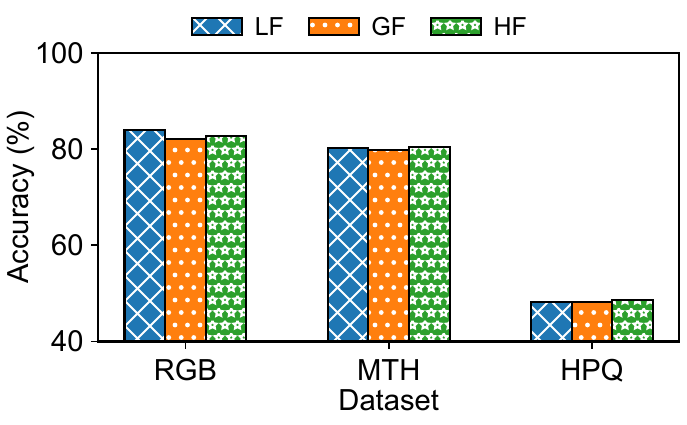}
    \vspace{-0.1in}
        \caption{Accuracy across feedback sources. LF, GF, and HF denote LLM, ground-truth (F1), and human feedback.}
        \label{fig:diff_feedback}
    \end{minipage}
    \hfill
    \begin{minipage}[t]{0.24\linewidth}
        \centering
        \includegraphics[width=0.95\linewidth]{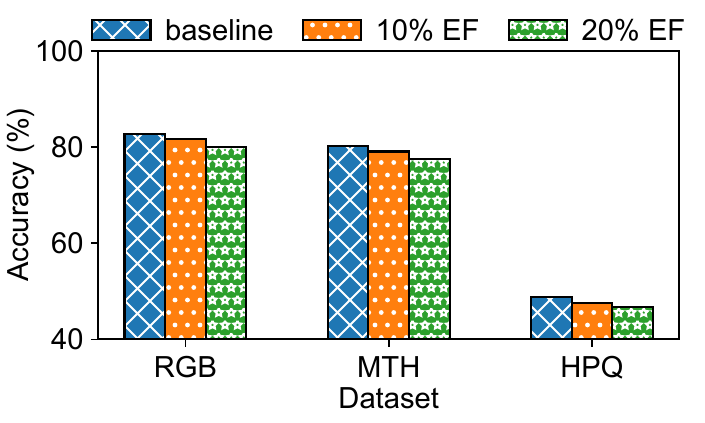}
    \vspace{-0.1in}
        \caption{Noise tolerance analysis. 
        $10\%$ and $20\%$ erroneous feedback (EF) is injected during each iteration.
        }
        \label{fig:robustness}
    \end{minipage}%
    \hfill
    \begin{minipage}[t]{0.24\linewidth}
        \centering
        \includegraphics[width=0.9\linewidth]{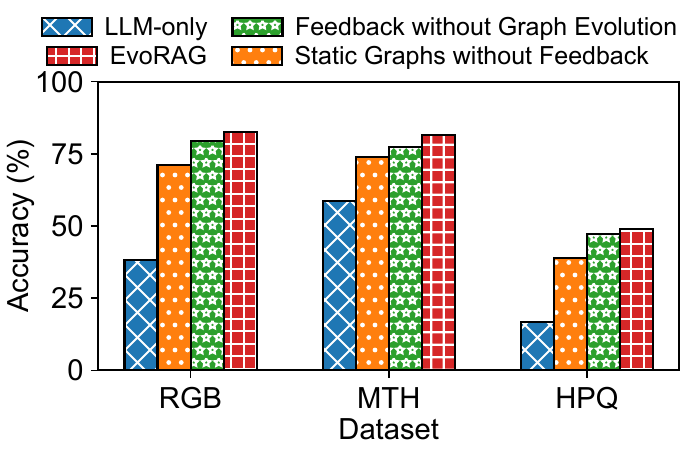}
    \vspace{-0.1in}
        \caption{Ablation study of feedback mechanism.}
        \label{fig:ablation_revision}
    \end{minipage}
    \hfill
    \begin{minipage}[t]{0.24\linewidth}
        \centering
        \includegraphics[width=0.95\linewidth]{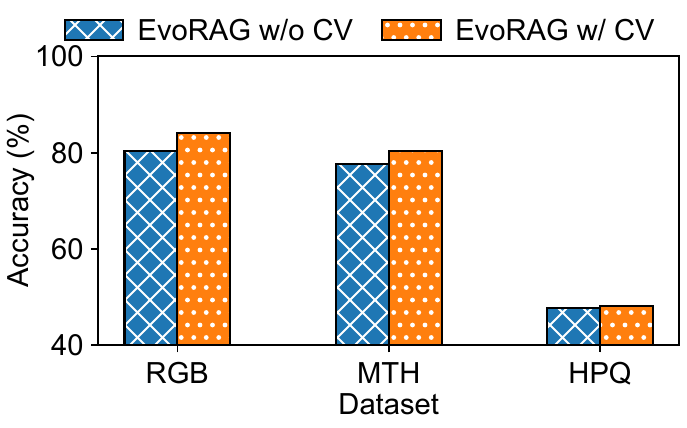}
    \vspace{-0.1in}
        \caption{Ablation study on the use of Fidelity and Conflict for cross-validation (CV).}
        \label{fig:vary_prompt}
    \end{minipage}
    \vspace{-0.3in}
\end{figure*}

\subsection{The Improvement of KG Quality}

KG evolution is performed once per iteration, where each iteration processes a set of training queries in batches, with contribution scores updated after each batch. 
We track training accuracy and the number of problematic triplets. Problematic triplets, defined in Section~\ref{sec:motivation} as irrelevant, outdated, incorrect, or long-path connections, are manually annotated.

Figure~\ref{fig:noise_degrad} shows that as iterations increase, the ratio of problematic triplets steadily decreases while accuracy improves and stabilizes after approximately $6$ iterations across all datasets. This indicates that \system can effectively incorporate useful feedback within a few iterations. On RGB, \system suppresses around $83.01\%$ of problematic triplets, achieving larger gains than HPQ due to the dataset’s high level of noisy facts. Performance improvements are most significant in early iterations, where \system can rapidly identify and down-weight misleading triplets. As iterations progress, the performance gradually converges as the system stabilizes.

\subsection{Impact of Different Feedback Sources}
\label{sec:diff_feedback}

To evaluate the applicability of \system, we compare the accuracy under different feedback sources. 
\fzb{Specifically, we consider LLM-generated feedback, ground-truth-based feedback (F1), and human feedback obtained by averaging independent five-point satisfaction ratings from ten Ph.D. students.}
As shown in Figure \ref{fig:diff_feedback}, \system achieves comparable performance across all settings, with an average difference of $0.75\%$.
This suggests that the performance gains mainly stem from the availability of feedback rather than their specific source.

\subsection{Feedback Tolerance Analysis}
\label{sec:tolerance}

To evaluate the impact of noisy feedback on \system, we simulate erroneous feedback by randomly selecting a fraction of the collected feedback and inverting the correctness. 
In the noise injection, we randomly select $10\%$ and $20\%$ of the feedback scores excluding neutral scores of 3, and flip them: scores $1-2$ are converted to $4-5$, and scores $4-5$ are converted to $1-2$. 
This process simulates imperfect or noisy feedback, reflecting realistic scenarios where evaluation judgments may be inaccurate or inconsistent.

The experimental results are reported in Figure \ref{fig:robustness}. We observe that \system remains tolerant under noisy feedback. Even with $10\%$ and $20\%$ erroneous feedback, accuracy drops only slightly, by $1.15\%$ and $2.43\%,$ respectively. 
This resilience stems from the cumulative update of contribution scores across queries and reasoning paths. Since each triplet is evaluated from diverse contexts, occasional erroneous feedback has a transient impact, while consistent signals dominate over time, enabling accurate convergence despite noise.

\subsection{Ablation Study}

\begin{figure}
    \centering
    \includegraphics[width=\linewidth]{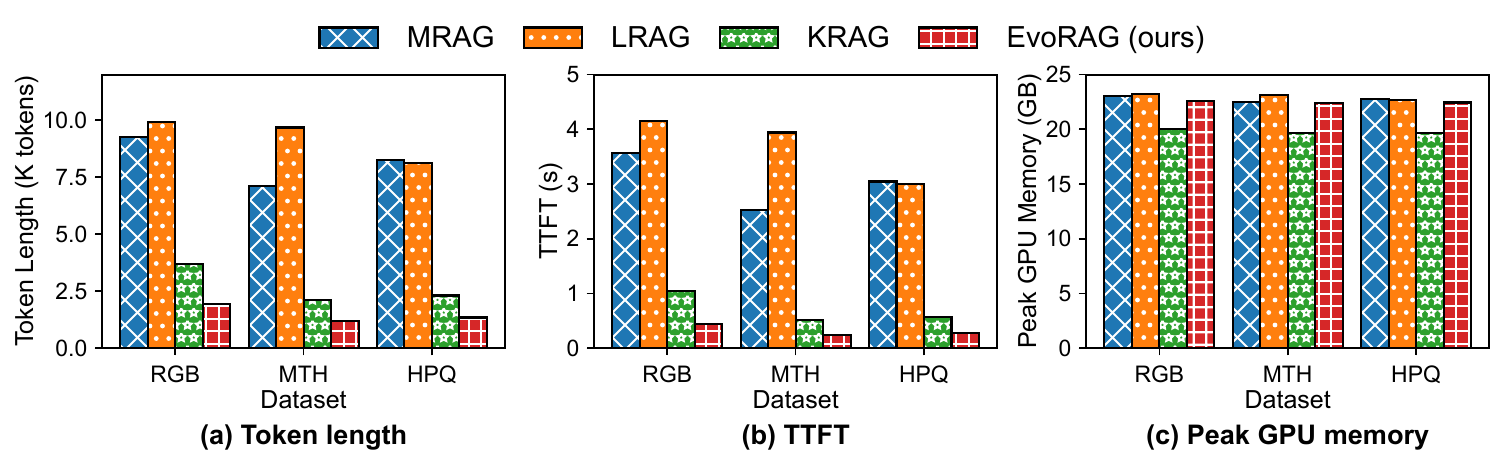}
    \vspace{-0.25in}
    \caption{\fzb{Comparison of \system with KG-RAG frameworks in terms of prompt length, TTFT, and peak GPU memory.}}
    \label{fig:token_cost}
    \vspace{-0.2in}
\end{figure}

\subsubsection{Ablation study of Feedback Mechanism}

To evaluate the contribution of the feedback mechanism, we compare four settings: LLM-only, LLM with static KG (SG), LLM+SG with feedback (FB), and LLM+SG+FB with graph evolution (i.e., \system).
As shown in Figure \ref{fig:ablation_revision}, SG outperforms the LLM-only by $23.42\%$ in accuracy, demonstrating KG-RAG’s ability to supplement missing knowledge. FB further improves accuracy by $6.6\%$ by prioritizing reliable triplets via feedback, while \system adds $3.07\%$ by updating the KG to prune low-utility triplets and reinforce correct long-range reasoning.
These results demonstrate the effectiveness of the feedback mechanism in improving reasoning accuracy.

\subsubsection{Ablation Study of Cross-validation Mechanism}

We evaluate path utility updates with and without Fidelity and Conflict. As shown in Figure~\ref{fig:vary_prompt}, incorporating these cross-validation constraints improves the average accuracy by $2.24\%$ over the variant without cross-validation, confirming their effectiveness in filtering unreliable or contradictory paths.

\subsection{\fzb{Efficiency and Resource Overhead}}

\fzb{\system uses updated contribution scores to suppress noisy triplets during retrieval. This reduces the input context and consequently lowers token cost and LLM inference latency. To quantify these benefits, we evaluate prompt length, time-to-first-token (TTFT), and peak GPU memory. As shown in Figure~\ref{fig:token_cost}, \system reduces the average prompt length by $4.6\times$ compared with MRAG, LRAG, and KRAG, resulting in consistently lower TTFT. MRAG and LRAG retrieve substantial associated text chunks, while KRAG lacks mechanisms to suppress semantically redundant or low-quality triplets. In contrast, \system retains task-relevant knowledge while filtering unnecessary context. Figure~\ref{fig:token_cost}(c) reports peak GPU memory usage. On average, \system peaks at $22.49$~GB of GPU memory, comparable to existing KG-RAG frameworks. Despite additional training, \system incurs limited memory overhead through shorter prompts and KV-cache reuse. Overall, \system improves retrieval efficiency and response latency while maintaining manageable resource consumption.}

\subsection{Runtime Performance Analysis}

\begin{figure}
    \centering
    \includegraphics[width=\linewidth]{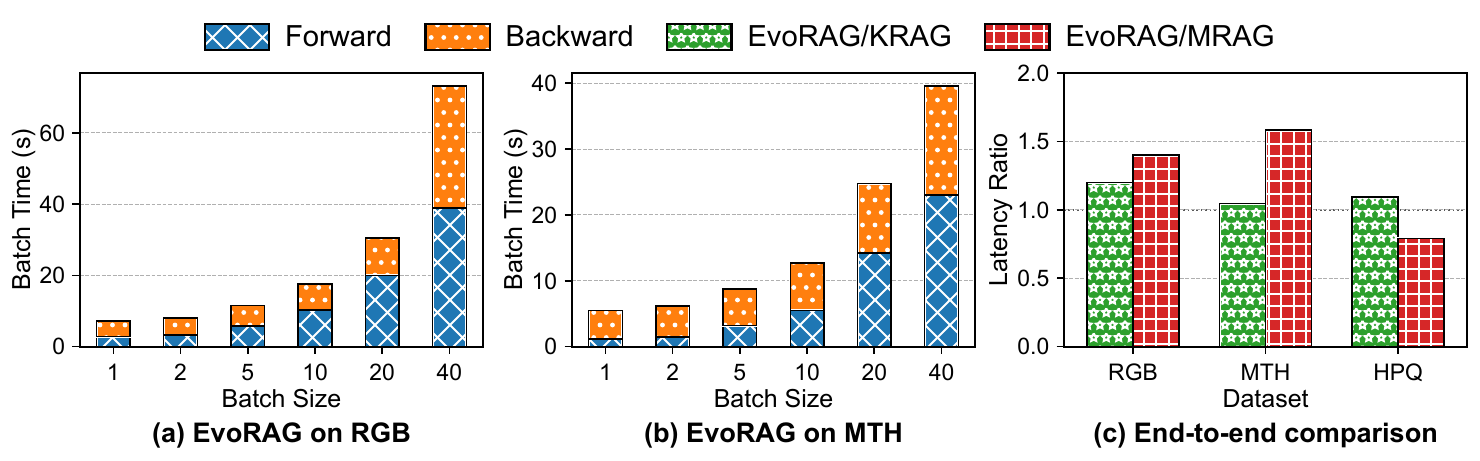}
    \vspace{-0.25in}
    \caption{\fzb{Runtime performance analysis of \system.}}
    \label{fig:performance}
    \vspace{-0.2in}
\end{figure}

\fzb{Feedback backpropagation introduces additional overhead compared to traditional KG-RAG frameworks. To reduce this overhead, we exploit the shared prefix between forward and backward prompts, which contains the same query and retrieved reasoning paths. We therefore adopt KV-cache reuse in vLLM \cite{vllm_sosp23}. We evaluate \system under different batch sizes. As shown in Figures \ref{fig:performance}(a) and (b), KV-cache reuse reduces the relative cost of backpropagation as the batch size increases. At a batch size of $20$, backpropagation accounts for only $23.90\%$ of the total runtime. Figure \ref{fig:performance}(c) further compares the end-to-end latency of \system with KRAG and MRAG. Across the three datasets, \system incurs approximately $1.12\times$ the latency of KRAG and $1.27\times$ that of MRAG on average. For HPQ, \system is faster than MRAG because MRAG constructs a substantially larger graph, resulting in higher retrieval latency. Overall, these results demonstrate that the overhead introduced by \system is well-controlled, making it suitable for deployment in real-world online scenarios.}

\subsection{Sensitivity Analysis of Entity and Path Numbers}
\label{sec:secsitive_study}

\begin{figure}
    \centering
    \includegraphics[width=0.8\linewidth]{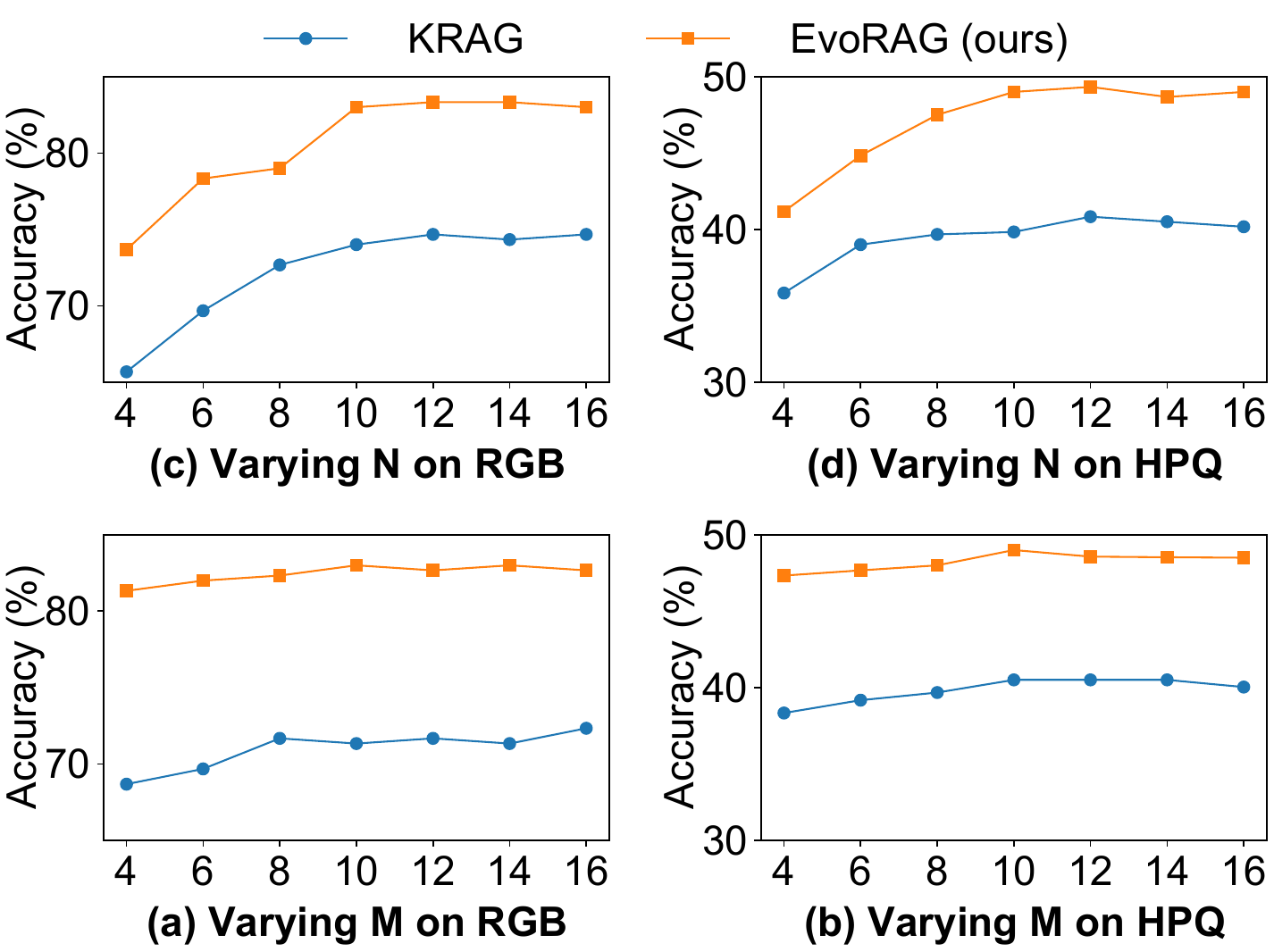}
    \vspace{-0.06in}
    \caption{Impact of hyperparameters on accuracy. $N$ indicates the number of retrieved entities; $M$ indicates the number of retrieval paths per entity.}
    \label{fig:Sensitive_study}
    \vspace{-0.15in}
\end{figure}

For KG-RAG frameworks, the number of entities ($N$) and reasoning paths per entity ($M$) directly impact the amount of knowledge retrieved and overall accuracy. We analyze the sensitivity of them by varying one parameter at a time while keeping the other fixed at $10$. The experimental results are shown in Figure \ref{fig:Sensitive_study}. We observe that increasing either $N$ or $M$ generally improves accuracy, as more retrieved knowledge provides a richer context for generation.
However, the gains diminish when $N$ and $M$ reach 10, where additional paths contribute marginally to performance, indicating that a limited number of paths are sufficient to enhance generation.
Notably, accuracy is more influenced by the number of entities than by the number of paths. This is because adding entities introduces more diverse and potentially relevant information.

\subsection{Scalability on Large Dataset}

\begin{figure}
    \centering
    \includegraphics[width=\linewidth]{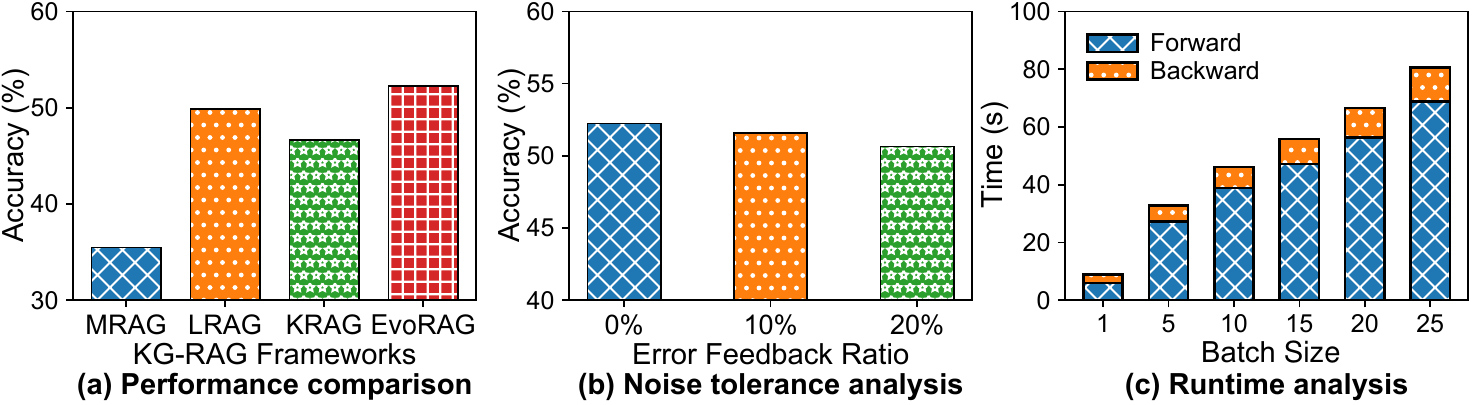}
    \vspace{-0.25in}
    \caption{Evaluation on 5,000 queries with large KG (890,389 entities, 1,043,360 triplets). (a) Accuracy comparison. (b) Noise tolerance analysis. (c) Performance analysis.}
    \label{fig:large_scale}
    \vspace{-0.2in}
\end{figure}

We evaluate \system in a large-scale setting with 5,000 queries sampled from HotpotQA on a KG containing 890,389 entities and 1,043,360 triplets. Figure~\ref{fig:large_scale} summarizes the results.
As shown in Figure~\ref{fig:large_scale}(a), \system improves the average accuracy over MRAG, LRAG, and KRAG by $8.25\%$, demonstrating that the feedback mechanism remains effective at scale.
Figure~\ref{fig:large_scale}(b) shows that each $10\%$ increase in erroneous feedback reduces accuracy by less than $1\%$, indicating strong noise tolerance due to cumulative aggregation across queries.
Figure~\ref{fig:large_scale}(c) shows that while the larger KG increases overall runtime (mainly in forward propagation due to retrieval overhead), backpropagation remains lightweight, accounting for only $15.35\%$ of total runtime at batch size 20.

\subsection{Effectiveness Analysis of Feedback Mechanism}

In this section, we analyze how feedback backpropagation affects the KG and retrieval results, and present a case study on representative queries.

\begin{table}[t]
    \centering
    \caption{Proportions of added and removed triplets in retrieval results after KG evolution.}
    \vspace{-0.06in}
    \small
    \begin{tabular}{c|c|c|c}
        \hline

        \hline
        \textbf{Proportion of Triplets} & \textbf{RGB} & \textbf{MTH} & \textbf{HPQ}\\
        \hline
        
        \textbf{Added} & 24.41\% & 21.06\% &13.46\% \\
        \textbf{Removed} & 28.97\% & 38.23\% & 17.31\% \\
        \hline

        \hline
    \end{tabular}
    \vspace{-0.1in}
    \label{tab:triplets_change}
\end{table}

\subsubsection{Changes in Retrieval Results}
We analyze the impact of KG evolution on retrieval by comparing the retrieved triplets before and after evolution. As shown in Table~\ref{tab:triplets_change}, $28.17\%$ of low-contribution triplets are removed, while $19.64\%$ of high-contribution triplets are newly introduced. This shows that KG evolution shifts retrieval away from selecting unhelpful relations and toward triplets that have proven useful.

\subsubsection{Case study}

\begin{figure}
    \centering
    \includegraphics[width=0.8\linewidth]{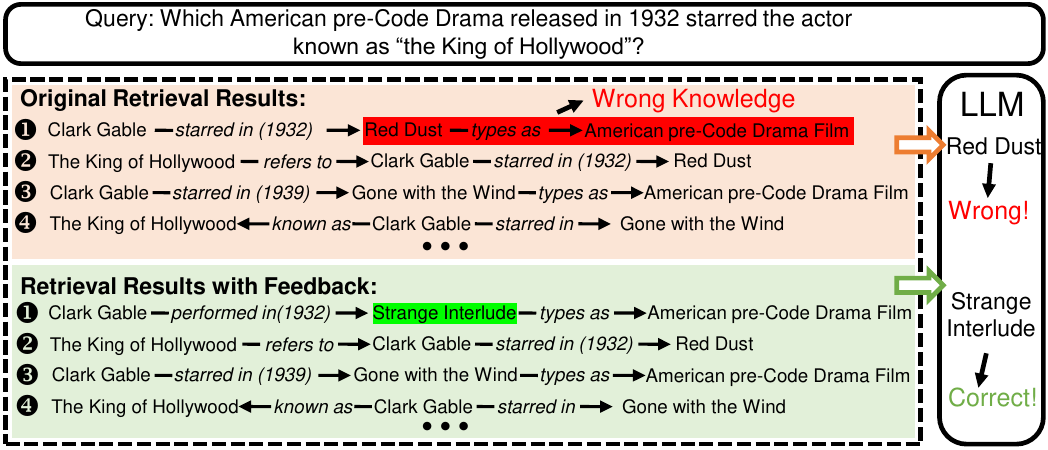}
    \vspace{-0.06in}
    \caption{Comparison of retrieved paths of a query over HPQ before and after 8 feedback iterations.}
    \label{fig:case_study}
    \vspace{-0.2in}
\end{figure}

Figure \ref{fig:case_study} shows a case study for a query from the HPQ dataset. In the original retrieval, path \ding{202} contains a clear factual error, but is ranked higher due to its high semantic similarity to the query, causing the correct path to Strange Interlude to be ranked lower. After feedback continuously indicates an incorrect answer in 8 iterations, \system suppresses this noisy triplet by reducing its contribution score and reinforces correct triplets. As a result, the correct path becomes more salient in retrieval, leading to a correct answer in subsequent queries.

\subsection{The Effectiveness of Feedback Backpropagation in MRAG and LRAG}

\begin{figure}[t]
    \centering
    \includegraphics[width=0.8\linewidth]{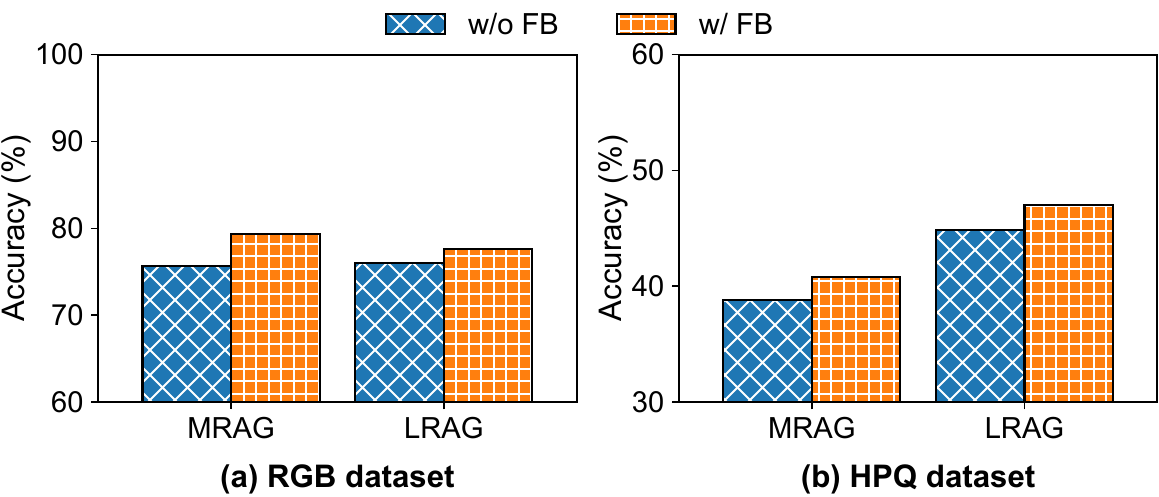}
    \vspace{-0.1in}
    \caption{Accuracy comparison of MRAG and LRAG on two datasets with Feedback Backpropagation (FB).}
    \vspace{-0.2in}
    \label{fig:kgimprove_MRAG_LRAG}
\end{figure}

\fzb{To evaluate the transferability of feedback backpropagation beyond KRAG, we apply the same mechanism to MRAG and LRAG by assigning contribution scores to retrieved text chunks.}
As shown in Figure~\ref{fig:kgimprove_MRAG_LRAG}, the integration improves the average accuracy of MRAG and LRAG by $2.83\%$ and $1.92\%$, respectively. 
The gains are smaller than those achieved by the triplet-level framework, as each text chunk aggregates multiple facts and thus provides coarser-grained feedback signals.

\subsection{Performance with Various LLM Backends}

\begin{figure}
    \centering
    \includegraphics[width=\linewidth]{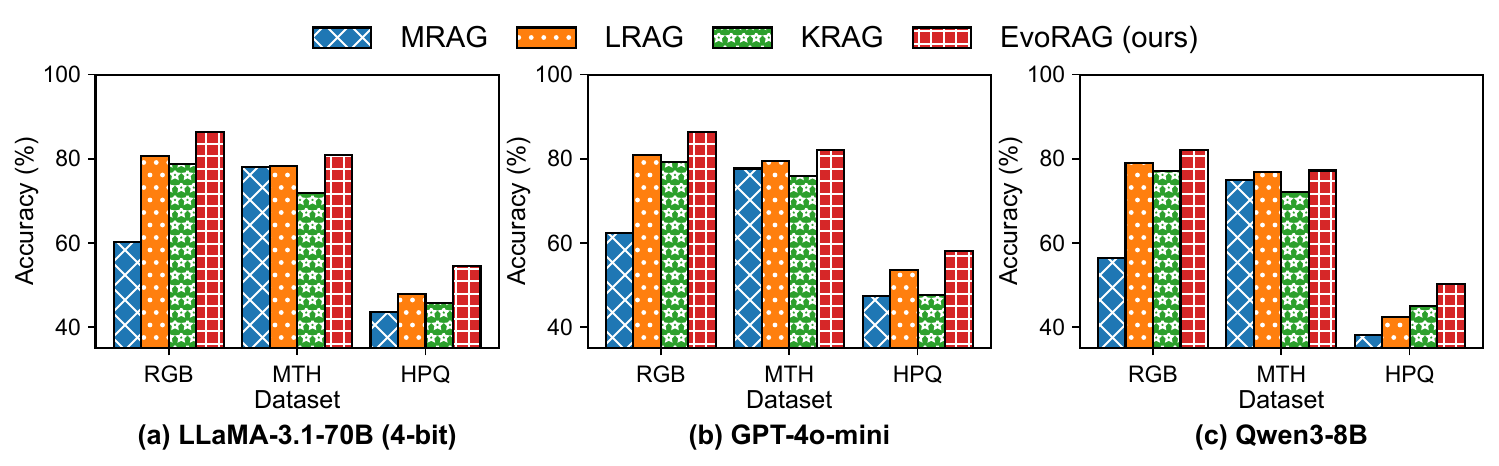}
    \vspace{-0.25in}
    \caption{\fzb{Accuracy comparison of different KG-RAG frameworks across different LLM backends.}}
    \vspace{-0.2in}
    \label{fig:vary_LLM}
\end{figure}

\fzb{We evaluate the generality of \system across different LLM backends by replacing Qwen2.5-32B with Llama-3.1-70B (4-bit), Qwen3-8B, and GPT-4o-mini for all LLM-dependent components in \system and the baselines. As shown in Figure~\ref{fig:vary_LLM}, \system outperforms existing RAG methods, improving average accuracy by $8.95\%$, $7.37\%$, and $8.31\%$ with Llama-3.1-70B, Qwen3-8B, and GPT-4o-mini, respectively. These results indicate that the performance gains of \system are robust to the choice of LLM backend.}

\begin{figure}[t]
    \centering
    \begin{minipage}[t]{0.44\linewidth}
        \centering
        \includegraphics[width=0.95\linewidth]{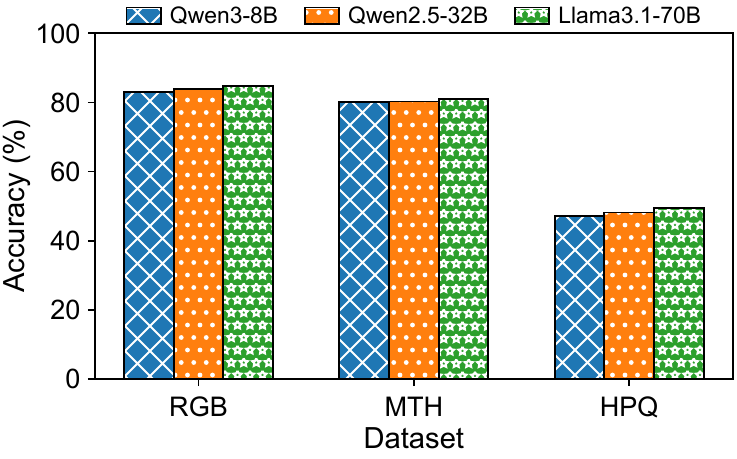}
        \vspace{-0.1in}
        \caption{Impact of different path evaluators.}
        \label{fig:path_evaluator}
    \end{minipage}%
    \hfill
    \begin{minipage}[t]{0.55\linewidth}
        \centering
        \includegraphics[width=0.95\linewidth]{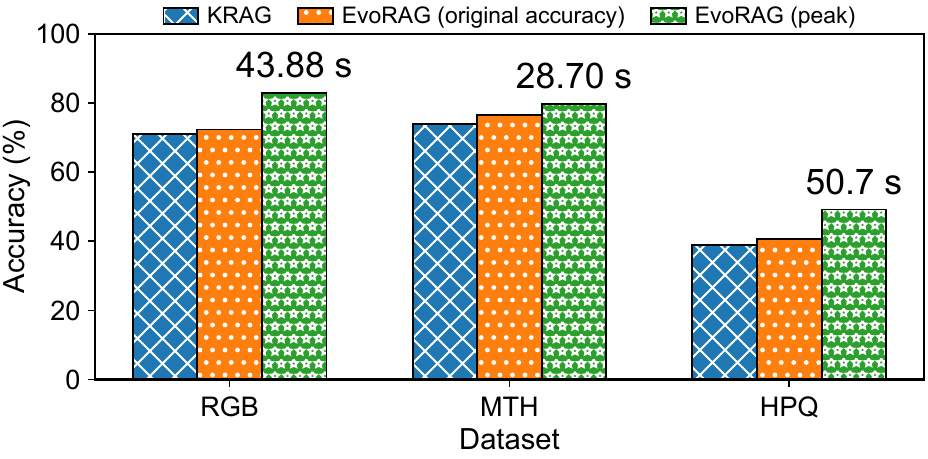}
        \vspace{-0.1in}
        \caption{Accuracy improvement of cold-start adaptation.}
        \label{fig:feedback_adaptation}
    \end{minipage}
    \vspace{-0.2in}
\end{figure}

\subsection{\fzb{Impact of the Path Evaluator}}
\fzb{
To examine the effect of the path evaluator, we fix the QA model to Qwen2.5-32B and vary only the path evaluator among Qwen3-8B, Qwen2.5-32B, and Llama-3.1-70B. Figure~\ref{fig:path_evaluator} shows average accuracy gains over KRAG of 8.72\%, 9.47\%, and 10.33\%, respectively, demonstrating that \system remains effective even with a smaller evaluator.
}

\subsection{\fzb{Cold-Start Analysis}}
\fzb{
To evaluate cold-start adaptation to unseen queries, we initialize \system without prior feedback. As shown in Figure~\ref{fig:feedback_adaptation}, feedback accumulation improves QA accuracy, reaching peak performance within 29-51 seconds across datasets. These results show that \system rapidly adapts to unseen queries and continually learns from accumulated feedback.
}

\section{Related Work}

\Paragraph{KG-RAG framework.} 
KG-RAG frameworks can be broadly grouped into two categories. One line of work represents knowledge as textual chunks and uses graph structures mainly for indexing inter-chunk relations~\cite{Raptor_ICLR24, PG-RAG_arxiv24, TOG2_arxiv24, Graphcoder_arxiv24, HippoRAG_NIPS24, MicrosoftGraphRAG_arxiv_2024, lightrag_arxiv_2024}. While effective for summarization, these methods are less suitable for complex reasoning.
Another line explicitly constructs and leverages KGs to support multi-hop reasoning over structured triplets~\cite{ProLLM_bioRxiv24, TOG_ICLR24, KGRAG_arxiv24, Knowgpt_NIPS24, KELP_ACL24, kag_arxiv24, PathRAG_arxiv25, DALK_EMNLP24, RuleRAG_arxiv24, GNNRAG_arxiv_2024, Graph_CoT_ACL24, Gretriever_NIPS24, Structrag_arxiv24, GraphReader_EMNLP24, StructureGuided_arxiv24}. However, most of these approaches emphasize retrieval and prompt design, without fully exploiting the KG’s capacity for complex reasoning, which is critical for accurate responses~\cite{GraphRAGBench_arxiv25}.

\Paragraph{KG refinement.}
KG refinement focuses on enhancing the factual accuracy and utility of KGs by removing redundant triplets, correcting incorrect facts, and adding missing information.
Rule-based methods \cite{Probabilistic_KG_noise_detection_CVPR19, Probabilistic_KG_noise_detection_CIKM19, KGRefine_survey_2016, rule_base_KG_noise_detection_PRICAI21} rely on logical constraints but scale poorly, while DL-based approaches \cite{mlkgr_survey_2024, embed_rulelearn_KGR_TKDE19, KGRsurvey_2021, TransE_NIPS13, TransT_PKDD17, TransG_ACL16, RotatR_ICLR19} typically operate offline and require high-quality training data. Recently, LLM-based methods \cite{LLMKGR_GenAIK25, KICGPT_EMNLP23, TCROF_arxiv25} assess semantic plausibility or generate facts, but remain detached from downstream tasks.

\Paragraph{KG validation.}
KG validation \cite{KGT_arxiv24, KGV_IPM25, cleangraph_arxiv24} assesses triplet correctness by relying on external evidence and expert-curated annotations, which often require task-specific pipelines and substantial expert involvement, thereby limiting scalability.

\Paragraph{Feedback-driven model optimization.}
These approaches \cite{ouyang2022training, yu2025self, zhang2024cppo} optimize model outputs in a feedback-driven manner via fine-tuning or reinforcement learning, yet operate solely at the model level without updating the KG.

\fzb{
Unlike these approaches, the novelty of \system lies in introducing feedback backpropagation: response feedback is propagated along reasoning paths to reusable triplets that guide retrieval and relation evolution.
}

\section{Conclusion}

We propose \system, a self-evolving KG-RAG framework that leverages real-time feedback to continuously refine the \fzb{static} KG and improve reasoning accuracy. \system introduces a feedback backpropagation mechanism that connects response-level feedback and triplet-level knowledge updates, which attributes feedback to individual reasoning paths and propagates it to adjust the scores of involved triplets. This establishes a closed loop, where reasoning feedback drives KG refinement, and the evolving KG improves future reasoning. The experimental results demonstrate that \system is both effective and robust, offering a scalable solution for adaptive and scalable KG evolution in real-world applications.

\section*{Acknowledgment}
This work was supported by the National Natural Science Foundation of China (U2241212, 62461146205, 62602108), the Distinguished Youth Foundation of Liaoning Province (2024021148-JH3/501), and Huawei. Yanfeng Zhang and Qiange Wang are the corresponding authors.

\section*{AI-Generated Content Acknowledgement}
The authors used ChatGPT to help polish the language of the manuscript. After using this tool, the authors carefully reviewed and edited all content as needed, and take full responsibility for the final content of the work.

\bibliographystyle{IEEEtran}
\bibliography{ref}

\end{document}